\documentclass[12pt,a4paper]{scrartcl}

\usepackage{LCVision}

\usepackage{LCVision_definitions}

\usepackage{mhchem}
\usepackage{multirow}
\usepackage{multicol}
\usepackage{threeparttable}

\usepackage{xpatch}
\usepackage{numprint}

\makeatletter
\xpatchcmd\@HepConStyle
 {\edef\@upcode{\updefault}}
 {\ifdefined\shapedefault\edef\@upcode{\shapedefault}\else\edef\@upcode{\updefault}\fi}
 {}{}
\makeatother

\RedeclareSectionCommand[
  beforeskip=1.\baselineskip,
  afterskip=.5\baselineskip]{subsection}
\RedeclareSectionCommand[
  beforeskip=1.\baselineskip,
  afterskip=.5\baselineskip]{subsubsection}

\usepackage{todonotes}
\usepackage{xpatch}
\makeatletter
\xpretocmd{\todo}{\@bsphack}{}{}
\xapptocmd{\todo}{\@esphack}{}{}
\makeatother

\usepackage{booktabs}
\usepackage{csquotes}

\usepackage{graphicx} 
\usepackage{appendix}

\def\beq{\begin{equation}}
\def\eeq#1{\label{#1}\end{equation}}
\def\eeqn{\end{equation}}

\newenvironment{Eqnarray}%
   {\arraycolsep 0.14em\begin{eqnarray}}{\end{eqnarray}}
\def\beqa{\begin{Eqnarray}}
\def\eeqa#1{\label{#1}\end{Eqnarray}}
\def\eeqan{\end{Eqnarray}}

\let\bar=\overbar

\def\lsim{\mathrel{\raise.3ex\hbox{$<$\kern-.75em\lower1ex\hbox{$\sim$}}}}
\def\gsim{\mathrel{\raise.3ex\hbox{$>$\kern-.75em\lower1ex\hbox{$\sim$}}}}

\def\del{\partial}
\def\Dslash{\not{\hbox{\kern-4pt $D$}}}
\def\dslash{\not{\hbox{\kern-2pt $\del$}}}

\def\Dlr{\mathrel{\raise1.5ex\hbox{$\leftrightarrow$\kern-1em\lower1.5ex\hbox{$D$}}}}

\def\Pl{{\mbox{\scriptsize Pl}}}

\def\ee{\ensuremath{\Pep\Pem}}

\def\msb{{\bar{\scriptsize M \kern -1pt S}}}

\def\drb{{\bar{\scriptsize D \kern -1pt R}}}

\usepackage[italic]{mathastext}

\DeclareCiteCommand{\citejournal}[\mkbibbrackets]
  {\usebibmacro{prenote}}
  {\usebibmacro{citeindex}%
   \printtext[bibhyperref]{\printfield{journaltitle}}%
   \iffieldundef{volume}
     {}%
     {\setunit{\addspace}%
     \printtext[bibhyperref]{\printfield{volume}}}%
   \setunit{\addspace}%
   \printtext[bibhyperref]{(\printdate)}%
   \iffieldundef{pages}
     {}
     {\setunit{\addspace}%
     \printtext[bibhyperref]{\printfield{pages}}%
     }%
     }
  {\multicitedelim}
  {\usebibmacro{postnote}}
  
\DeclareCiteCommand{\citesubmit}[\mkbibbrackets]
  {\usebibmacro{prenote}}
  {\usebibmacro{citeindex}%
   \printtext[bibhyperref]{\printfield{journaltitle}}%
   \setunit{\addspace}%
   \printtext[bibhyperref]{(\printdate)}}
  {\multicitedelim}
  {\usebibmacro{postnote}}
  
  \DeclareCiteCommand{\citeconf}[\mkbibbrackets]
  {\usebibmacro{prenote}}
  {\usebibmacro{citeindex}%
   \printtext[bibhyperref]{\printfield{howpublished}}%
   \setunit{\addspace}%
   \printtext[bibhyperref]{(\printdate)}}
  {\multicitedelim}
  {\usebibmacro{postnote}}

\title{The Linear Collider Facility (LCF) at CERN\newline
{\footnotesize updated version May 26, 2025} }

\nolicence 

\notitlestamp 

\date{\today}

\abstract{\noindent
In this paper we outline a proposal for a Linear Collider Facility as the next flagship project for CERN. 
This proposal offers the opportunity for a timely, cost-effective and staged construction of a new collider that will be able to comprehensively map the Higgs boson's properties, including the Higgs field potential, thanks to a large span in centre-of-mass energies and polarised beams.
A comprehensive programme to study the Higgs boson and its closest relatives with high precision requires data at centre-of-mass energies from the \PZ pole to at least \SI{1}{\TeV}. It should include measurements of the Higgs boson in both major production mechanisms, $\ee\to \PZ\PH$ (Higgs-strahlung) and $\ee\to \PGn \PAGn \PH$ ($\PW\PW$ fusion), precision measurements of gauge boson interactions as well as of the $\PW$ boson, Higgs boson and top-quark masses, measurement of the top-quark Yukawa coupling through $\ee \to \PQt\PAQt\PH$, measurement of the Higgs boson self-coupling through $\PH\PH$ production, and precision measurements of the electroweak couplings of the top quark. In addition, $\ee$ collisions offer discovery potential for new particles complementary to HL-LHC.
The facility we propose robustly satisfies these scientific goals.
With a total length of \SI{33.5}{km}, two interaction regions as well as additional R\&D and fixed-target experiments, it offers significant flexibility to take into account scientific and strategic developments. 
From today's perspective, we propose to equip the Linear Collider Facility in a first stage with superconducting RF cavities for polarised $\ee$ collisions at a centre-of-mass energy of \SI{250}{GeV} with a luminosity of $2.7\times10^{34}$\,cm$^{-2}$s$^{-1}$, which requires an investment of about \SI{8.3}{BCHF}. 
With a preparatory phase of eight years, followed by ten years of construction starting earliest in 2034, this first stage could start beam commissioning and data-taking a decade later.
First upgrades comprise doubling of the luminosity for \SI{0.8}{BCHF} and an increase of energy up to at least \SI{550}{GeV}, which can be achieved with the same accelerator technology for about \SI{5.5}{BCHF}.
Later stages will involve further increase of luminosity and energy as well as other new capabilities that will further enhance the Higgs programme and extend the discovery potential for new physics. 
These upgrades will primarily be accomplished by accelerator technology innovations rather than by additional civil construction.

}

\addauthor{
\vspace*{1cm}

C.~Balazs\\
\emph{Monash University, Melbourne 3800 Victoria, Australia}
}

\addauthor{
G.~Taylor\\
\emph{University of Melbourne, Victoria 3010, Australia}
}

\addauthor{
W.~Mitaroff\\
\emph{Institute of High Energy Physics, Austrian Academy of Sciences, 1050 Vienna, Austria}
}

\addauthor{
A.H.~Hoang,
S.~Pl\"{a}tzer\\
\emph{University of Vienna, Faculty of Physics, 1090 Vienna, Austria}
}

\addauthor{
M.~Tytgat\\
\emph{Inter-University Institute for High Energies, Vrije Universiteit Brussel, 1050 Ixelles, Belgium}
}

\addauthor{
C.~Hensel\\
\emph{Centro Brasileiro de Pesquisas F\'isicas, Rio de Janeiro 22290-180, Brazil}
}

\addauthor{
A.B.~Bellerive\\
\emph{Carleton University, Ottawa, Ontario K1S5B6, Canada}
}

\addauthor{
F.~Corriveau\\
\emph{Centre for High-Energy Physics, McGill University, Montreal, Quebec H3A2T8, Canada}
}

\addauthor{
D.~Tuckler\\
\emph{TRIUMF, Wesbrook Mall, Vancouver, Canada}
}

\addauthor{
C.D.~Fu,
Q.~Ouyang,
H.R.~Qi,
M.Q.~Ruan\\
\emph{Institute of High Energy Physics, Chinese Academy of Sciences, Beijing, China}
}

\addauthor{
I.P.~Ivanov\\
\emph{Sun Yat-sen University, Guangzhou, Guangdong, China}
}

\addauthor{
M.~Ouchemhou,
T.~Robens\\
\emph{Rudjer Boskovic Institute, 10000 Zagreb, Croatia}
}

\addauthor{
J.~Cvach,
J.~Kvasni\v{c}ka,
I.~Pol\'{a}k,
J.~Z\'{a}le\v{s}\'{a}k\\
\emph{FZU - Institute of Physics of the Czech Academy of Sciences, 18200 Prague 8, Czechia}
% ...
}

\addauthor{
J.D.~Wells\\
\emph{Physics Dept, Ann Arbor, MI 48109, USA}
}

\addauthor{
H.~Baer\\
\emph{University of Oklahoma, Norman, OK 73019, USA}
}

\addauthor{
J.E.~Brau,
C.T.~Potter,
J.~Strube\\
\emph{University of Oregon, Eugene, OR 97403, USA}
}

\addauthor{
K.~Sugizaki\\
\emph{University of Pennsylvania, Philadelphia, PA 19104, USA}
}

\addauthor{
A.P.~White\\
\emph{University of Texas at Arlington, Arlington, TX 76019, USA}
}

\addauthor{
Y.~Bai,
S.~Dasu\\
\emph{University of Wisconsin--Madison, Madison, WI 53706, USA}
}

\addauthor{
\begin{flushleft}
{$^{0}$}Also at: King Abdulaziz City for Science and Technology (KACST), Riyadh, Saudi Arabia
{$^{1}$}Also at: NSC KIPT National Science Center "Kharkiv Institute of Physics and Technology" of The National Academy of Sciences of Ukraine, Kharkiv, Ukraine\\
{$^{2}$}Also at: University of Hamburg\\
{$^{3}$}Also at: Institute for Theoretical Physics I, Heinrich Heine University, D\"{u}sseldorf\\
{$^{4}$}Also at: St. Joseph's University, Bangalore 560027, India\\
{$^{5}$}Also at: Indian Institute of Science Education and Research Kolkata, Mohanpur, West Bengal, India\\
{$^{6}$}Also at: Nambu Yoichiro Institute of Theoretical and Experimental Physics (NITEP)\\
{$^{7}$}Also at: ICEPP\\
{$^{8}$}Also at: Kavli IPMU\\
{$^{9}$}Also at: Ludwig-Maximilians-Universit\"at M\"unchen, 85748 Garching, Germany\\
{$^{10}$}Also at: Deutsches Elektronen-Synchrotron DESY, Germany\\
{$^{11}$}Also at: CERN, Geneva, Switzerland\\
{$^{12}$}Also at: Physics and Astronomy, Stony Brook University, NY 11794, USA\\
{$^{13}$}Also at: Argonne National Laboratory, Lemont, IL 60439, USA
\end{flushleft}
}

\graphicspath{ {./logos/}{./figures/} }

\DeclareTOCStyleEntry[beforeskip=.2cm]{section}{section}

\begin{document}

\titlepage
\pagenumbering{arabic}\setcounter{page}{1}



\section{Introduction}

The aim of the current update of the European Strategy for Particle Physics  (ESPP)  is to develop a ``visionary and concrete plan'' for the realisation of the next flagship project at CERN~\cite{ESPPremit}.  
In this report, we propose that the next CERN flagship project after the High Luminosity LHC should be a linear $\ee$ collider spanning centre-of-mass energies from the $\PZ$ pole to at least \SI{1}{\TeV}. 
We will give a cost and timeline for the initial stage of this project and describe options for later stages that encompass high-energy running and flexible responses to possible contingencies.

A linear collider, in its run at \SI{250}{\GeV} in the centre of mass, will already bring the precision on Higgs boson couplings, both to Standard Model and potential exotic particles appearing in its decay, to the percent level.  
At higher energies, it can provide precision measurements of the top quark mass, the top quark Yukawa coupling and the Higgs self-coupling. 
Also, it will make precision measurements of the electroweak couplings of the top quark relevant to the physics of its mass generation. 
A linear collider offers a high degree  of flexibility due to its intrinsic upgradability. 
The scope of its initial stage and the nature and timeline of upgrades can be chosen and adjusted throughout the project taking into account scientific developments and competition, availability of resources and global cooperation.
The previous European Strategy report stated, ``An electron-positron Higgs factory is the highest-priority next collider''~\cite{ESPP2020}.  
A linear collider is a true Higgs factory, capable of carrying out a comprehensive suite of measurements relevant to the Higgs boson that we have just described and that  we will quantify below.

CERN has the resources and experience to host this linear collider programme. 
With CERN's experience in building and operating the LHC, the laboratory is well prepared to lead an ambitious programme of linear collider physics, one that encompasses a complete exploration of the Higgs boson's properties.  
We present a broad project, with high 
energy reach, with two interaction regions allowing complementary detectors and optionally a rich beyond-collider programme.  
This machine allows that close attention can be paid to the needs
of physics, including the flexibility to respond to possible unexpected outcomes of the HL-LHC -- and of the first stage of the linear collider itself. 
Whatever the circumstances over the next decade and a half, this program will produce physics by mid 2040s, with results continuously expanding into the future.

This paper is organised as follows:  In Section~\ref{sec:phys}, we will discuss the
needs of a comprehensive programme to measure the properties of the Higgs
boson with high precision. This programme includes running at energies well above the top quark threshold to measure the crucial observables of top quark and Higgs boson pair production.
In Section~\ref{sec:baseline}, we will present our proposal
for a Linear Collider Facility at CERN.  This is a staged plan to maximise the flexibility of the proposal, allowing to respond to future physics discoveries and technology advancements. The initial stage is a linear $\ee$ collider and beam delivery region of length \SI{20.5}{km} in a tunnel of total length \SI{33.5}{km}, with two interaction regions sharing luminosity.  The cost 
of the initial stage is \SI{8}{BCHF}. The initial centre of mass energy would be \SI{250}{GeV}, but upgrades to the accelerator would reach \SI{550}{GeV}, \SI{1}{TeV}, or as high as \SI{3}TeV, depending on the technology and the physics objectives.
In Section~\ref{sec:upgrades}, we will describe paths for these energy upgrades, any of which will fulfil the comprehensive Higgs program outlined in Section~\ref{sec:phys}. We will also present additional upgrade paths
to provide dramatically increased luminosity, energy reach, or new physics capabilities -- addressing the long-term goal of a \SI{10}{TeV} parton centre-of-mass collider.  
In Section~\ref{sec:next} we will outline the next steps towards the realisation of the Linear Collider Facility.  We will give our conclusions in Section~\ref{sec:conclusions}.


\section{Precision study of the Higgs boson}
\label{sec:phys}

The study of the Higgs boson has been given priority in all recent
planning studies from Europe, the US, and Japan.
There is a good reason for this. The gauge
couplings of the Standard Model (SM) are now well understood and
well-tested. But many questions in particle physics remain
unanswered.  These include the origin of electroweak symmetry
breaking, the hierarchy of quark and lepton masses, the origin of CP
violation, and the origin of neutrino masses. In the SM,
all of these are accounted for by terms involving the Higgs boson whose
coefficients are adjusted by hand.  The true explanations must
involve new interactions and particles that couple to the Higgs boson.

For more than 40 years, particle physicists have been searching for new physics beyond the SM, but no convincing evidence has yet been found. The most promising approaches are now at the energy frontier, where the LHC experiments have carried out extensive new particle searches and have made great progress in understanding the Higgs boson.   Still, much opportunity for discovery is still available  here.  This motivates a study of the Higgs boson that is as precise and as complete as possible.

It is well appreciated now that studying the interactions of the Higgs
boson at an $\ee$  collider provides the best route to high-precision
measurements of the Higgs properties.  
The required programme has been enunciated in many places,  in particular in~\cite{ILC:2013jhg, Roloff:2018dqu} and, more recently, in~\cite{ILCInternationalDevelopmentTeam:2022izu} and~\cite{LCVision-Generic}. 
The programme requires several stages, at different $\ee$ centre-of-mass energies.

\begin{figure}[htb]
\begin{center}
  \begin{subfigure}{.5\textwidth}
    \centering
    \includegraphics[width=0.90\hsize]{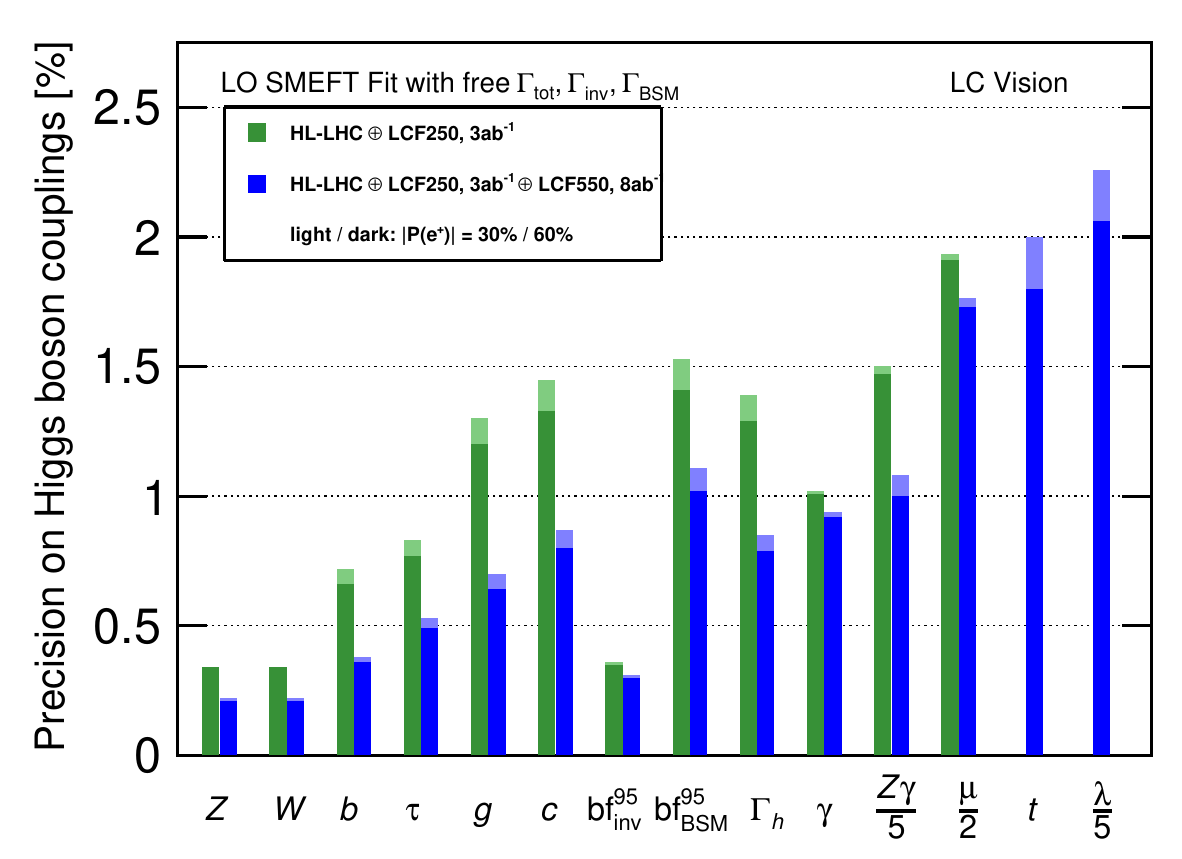}
    \caption{}
    \label{fig:projections:coup}    
  \end{subfigure}\hfill%
  \begin{subfigure}{.5\textwidth}
    \centering
    \includegraphics[width=0.90\hsize]{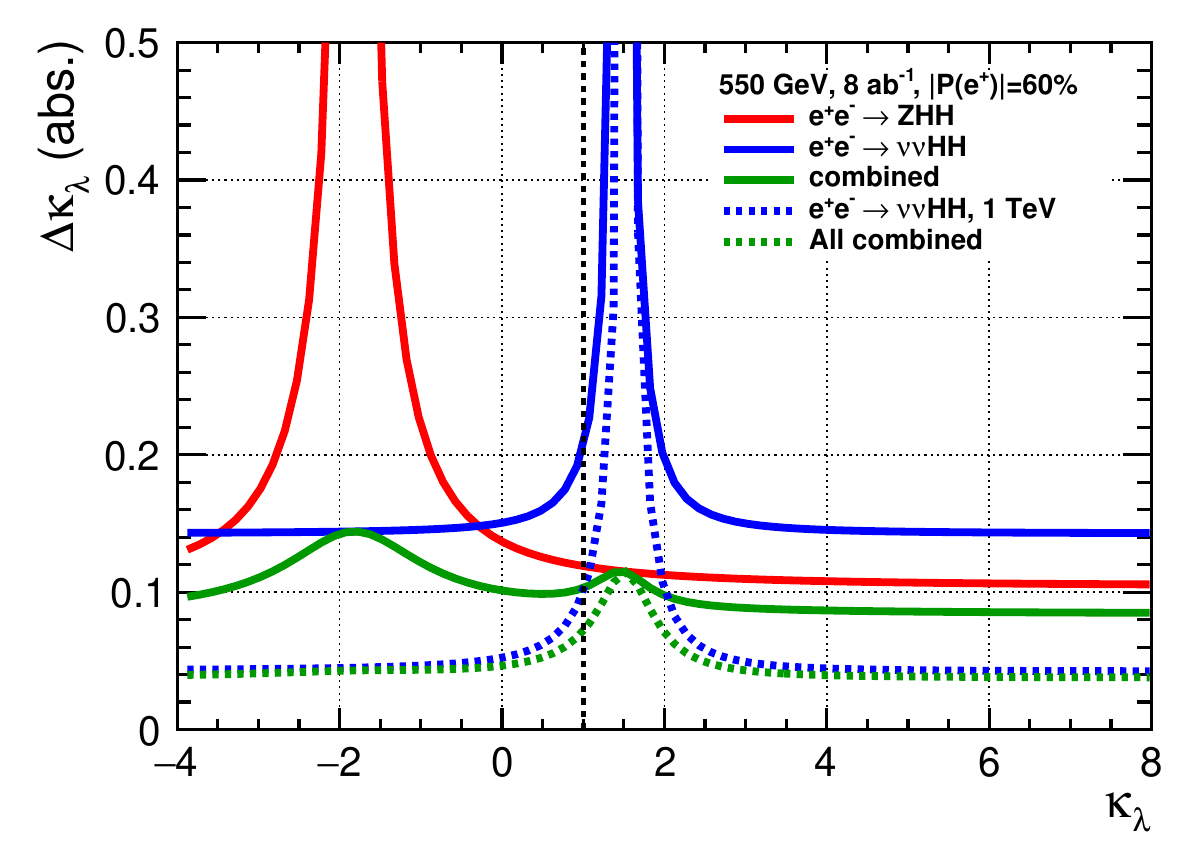}
    \caption{}
    \label{fig:projections:kala}    
  \end{subfigure}\hfill%

\caption{\label{fig:projections} Projected uncertainties on Higgs boson couplings corresponding to the run plan in Fig.~\ref{fig:runplan}. The methodology used to make these projections and projections for other choices of the run plan are presented in \cite{LCVision-Generic}. (a) Overview on various Higgs couplings; the bars for the muon Yukawa and the trilinear Higgs self-coupling $\lambda$ have been scaled by $1/2$ and $1/5$, respectively. (b) Absolute precision on trilinear Higgs self-coupling scaled to its SM value ($\kappa_{\lambda}$) as function of $\kappa_{\lambda}$ from the $\PZ\PH\PH$ and $\PW\PW$ fusion processes at \SI{550}{GeV} and \SI{1}{TeV}.}

\end{center}
\end{figure}

\begin{enumerate}
  \item  An initial stage at \SI{250}{\GeV}, recording a large data set of
    reactions $\ee\to \PZ\PH$. This will measure the Higgs branching
    fractions to \SI{1}{\%}  precision, measure the $\PZ\PH$ total
    cross section to \SI{1}{\%} precision to determine the absolute normalisation
    of Higgs couplings, measure the Higgs boson mass to $10^{-4}$
    precision, and search for non-standard Higgs boson decays, including invisible ones, to the
    $10^{-4}$ level of branching fractions. The expected precisions on the individual Higgs couplings from collecting \SI{3}{\abinv} of polarised data are shown as green bars in Fig.~\ref{fig:projections:coup}.

    This data set will also give the most precise measurements of the $\PW$ boson non-linear interactions and of electroweak fermion pair production, providing further opportunities for discovery. The $\PW$ boson's mass can be measured to about \SI{2}{MeV} using the \SI{250}{GeV} data. An optional one-year threshold scan could improve this to \SI{1.4}{MeV}.
 Also, at this stage, one can collect a few $10^9$ events on the $\PZ$ pole with polarised beams, to test the SM electroweak sector with high precision. In this regard, the use of beam polarisation  compensates almost three orders of magnitude in integrated luminosity with respect to unpolarised colliders.
 
  \item A second stage at \SI{550}{\GeV}, where the $\PW\PW$ fusion reaction $\ee\to \PGn \PAGn \PH$ has become the dominant production mechanism for Higgs bosons. 
  This will provide an independent setting for the measurement of Higgs branching fractions and searches for exotic decays.  
  It also allows the measurement of the top quark Yukawa coupling and the trilinear Higgs self-coupling through the processes $\ee\to \PQt\PAQt\PH$ and $\ee\to  \PZ\PH\PH$, respectively. 
  The blue bars in Fig.~\ref{fig:projections:coup} show the expected gain in precision on the individual Higgs couplings from \SI{8}{\abinv} of polarised data at \SI{550}{\GeV}, including a precision of \SI{2}{\%} or better on the top Yukawa and  of \SI{11}{\%} or better on the Higgs self-coupling (SM case). 
    Figure~\ref{fig:projections:kala} illustrates that thanks to the two complementary di-Higgs production processes, the expected combined precision on the self-coupling is nearly independent of the actual value nature might have chosen. 

    The top quark is the SM particle most strongly
    coupled to the Higgs boson and thus most likely  to be affected by
    new Higgs interactions.  This stage, well above the top quark
    threshold, will also provide precision measurements of the top quark electroweak couplings at the level of \SI{1}{\%} or better, and include a short run around the top pair production 
    threshold for a precise mass determination. 

    \item A third stage at at least \SI{1}{\TeV}, where the reaction of $\PW\PW$
      fusion to $\PH\PH$ is the dominant Higgs pair production
      mechanism.   This energy stage thus provides a second method to
      determine the trilinear Higgs self-coupling, and can put constraints on the quartic Higgs self-coupling~\cite{Stylianou:2023xit}. 
      The two Higgs pair production processes, $\PZ\PH\PH$ and $\PW\PW$ fusion, are strikingly complementary (see also Fig.~\ref{fig:projections:kala}); as the self-coupling is increased from its SM value, the $\PZ\PH\PH$ cross section increases while the $\PW\PW$ fusion cross section decreases.  Observing these deviations in the same experiment would give the most persuasive evidence for a non-standard value of the Higgs self-coupling.
      
      Anomalies in top quark or multi gauge boson couplings suggested in earlier stages will be
      dramatically larger, proportionally to $E_{CM}^2$ in typical cases.
      The ample production of (single) Higgs bosons in $\PW\PW$ fusion and $\PQt\PAQt\PH$ will allow us to further scrutinise the couplings of the Higgs boson to SM particles including the top quark and the $\PW$ boson.
    
    \end{enumerate}

 \noindent Any of these stages, but increasingly so the higher-energy ones, provide opportunities for discoveries via largely loophole-free tests of new particle existence to mass scales approaching the kinematic limit.
A linear $\ee$ collider can provide all these stages and, as we will discuss below, could be upgraded to energies of up to \SI{3}{\TeV}. This would allow precision studies of heavy Higgs bosons and other electroweak states discovered at the HL-LHC, or at the linear collider itself.

Our understanding of the Higgs boson and top quark couplings will be
best informed through global fitting of the entire data set to specific UV-complete extensions of the SM or -- as long as there is no hint which extension of the SM could be realised in nature --
to Effective Field Theories like e.g.\ Standard Model Effective Field Theory (SMEFT).  Such fits, and in particular the SMEFT fit, require
even better knowledge of precision electroweak parameters, the
$\PW$ and 
top quark masses, and the cross sections for 
$\ee\to \PWp\PWm$ and other electroweak reactions.  That will naturally
be achieved in the course of  this programme, with the additional brief runs
at the $\PZ$ pole and the $\PQt\PAQt$ threshold. Because the SMEFT
Lagrangian contains chiral operators, global SMEFT fits typically have degeneracies involving left- vs.\ right-handed couplings.  At a linear collider, the availability of
$\Pem$ and $\Pep$ beam polarisation will resolve these degeneracies, to  bring out the maximum information available from the data.

A full discussion of the linear collider physics programme, with descriptions of all of the above measurements, their combination through SMEFT fitting, and estimates of expected precision as a function of integrated luminosity, can be found in~\cite{LCVision-Generic}. 
Other, more extensive, discussions of the Higgs factory physics programme, can be found in earlier papers~\cite{ILCInternationalDevelopmentTeam:2022izu, Bambade:2019fyw, Moortgat-Pick:2015lbx, ILC:2013jhg, Moortgat-Pick:2005jsx}.

There is one more important point about this program that we would like to especially emphasise.   
Though there are windows for the discovery of new particles at the HL-LHC, it may be that our only opportunity to discover new physics with accelerators using current technologies will be through the discovery of deviations from the SM in precision measurements.   
Such a discovery will need to overcome a very substantial burden of proof. 
Deviations at the 3$\sigma$ level are common in LHC analyses, yet are not sufficient to question the SM.
The 7$\sigma$ deviation in the $\PW$ boson mass reported by the CDF collaboration~\cite{CDF:2022hxs} has been greeted by our community with great scepticism.  
A discovery via precision measurements will require not only a high level of precision, including control and detailed understanding of systematic uncertainties, but also the independent observation of the anomaly in different settings with different experimental challenges. 
Beam polarisation allows non-trivial cross-checks on any observed deviation. 
More importantly, the qualitative evolution of $\ee$ physics with centre-of-mass energy will provide many opportunities for confirmation of discoveries. 
In the programme that we have outlined above, the observation of an anomaly in Higgs boson couplings observed at \SI{250}{\GeV} could be confirmed by observation of the same anomaly in $\PW\PW$ fusion. 
A deviation from the SM in the Higgs self-coupling or the top quark Yukawa coupling observed at \SI{550}{\GeV} could be confirmed at the TeV stage through data from new processes. 
A precision Higgs programme that does not have this option invites confusion and stalemate. 
A multi-stage linear collider programme has the robustness needed to cement a true discovery.


\section{The Linear Collider Facility at CERN (up to about 550\,GeV)}
\label{sec:baseline}

The multi-stage $\ee$ programme described in the previous section would be carried out step by step, ideally with each step affordable within the current CERN budget. 
What is needed today is a detailed plan for the first stage, for which  construction could begin around 2034 and be completed a decade later as the HL-LHC programme is completed (c.f.\ Sec.~\ref{sec:next} and~\cite{LCF:EPPSU:backup} for more information on the timeline).

The timing here is important, especially for young scientists who will be key to both the HL-LHC program and the future Higgs factory. 
Prolonged uncertainty or delays in decision-making risk discouraging Early Career Researchers, leading to a loss of talent and expertise that could undermine the vitality of collider physics. 
Ensuring a clear and timely transition from HL-LHC to the next collider project will provide young scientists with long-term research opportunities, allowing them to contribute to both late-stage HL-LHC analyses and the design and construction of future detectors. 
The plan we propose aligns with the priorities identified in the European Early Career Researcher strategy submission~\cite{Arling:2025tah}, which emphasises technological innovation, alongside an ambitious baseline physics programme, as the driving factor in deciding CERN's next flagship project. 
Additionally, the lower priority~\cite{Arling:2025tah} places on a predefined upgrade path aligns with flexibility to shape the LCF's upgrade strategy based on future developments.

To provide for this, the first-stage linear collider should be based on well-understood technology and planning. Our proposal builds upon decades of work on the most mature linear collider concepts ILC~\cite{ILC-EPPSU:2025} and CLIC~\cite{CLIC-EPPSU:2025}, for which technical design reports, project implementation plans etc.\ have been presented~\cite{Adolphsen:2013kya, Aicheler:2018arh}. 
The ILC technical design has been under review for more than 10 years by ICFA and by several Japanese government committees. 
The costs given in this document have been obtained in the following way, as detailed in~\cite{LCF:EPPSU:backup}: 
\begin{itemize}
    \item {\bfseries{Accelerator costs:}} The costs for everything but the civil construction have been derived from the 2024 cost update for the ILC in Japan, which was based to a significant extent on new quotes from industry and reviewed by an international expert committee~\cite{ILC-EPPSU:2025}. 
    \item {\bfseries{Civil engineering and conventional systems:}} The civil engineering required at CERN is very similar to that needed for the first and second stages of the CLIC project, and has been costed in 2025 with unit prices in agreement with other CERN-hosted projects. 
    The conventional systems have been taken from the TDR estimate and escalated to 2024.   
    \item {\bfseries{Operation costs:}} The annual operating costs have been evaluated in 2025 according to common standards for CERN projects, e.g.\ CLIC. 
\end{itemize}

\paragraph{Superconducting RF as initial technology} Aiming for a minimal time to physics, we see a strategic advantage in choosing an accelerator based on superconducting radio-frequency cavities (SCRF) for the first stage. 
By now, many SCRF-based accelerators are being built or successfully operated (e.g.\ Eu.XFEL, LCLS-II, SHINE) and the corresponding technological expertise and industrialisation is available in laboratories and companies around the world. In addition, SCRF technology has seen important developments towards higher gradients and higher quality factors (for a recent summary of the state-of-the-art c.f.~\cite{LCVision-Generic}). 
In particular, a factor of two improvement in quality factors over the specification chosen about 15 years ago for ILC is now reliably achievable with a small modification of the cavity production recipe. 
The alternative possibility of starting with warm technology is discussed in the CLIC submission to the ESPPU~\cite{CLIC-EPPSU:2025,Adli:ESU25RDR}. 

We envision that succeeding stages of the programme will bring in new accelerator technologies that are now under development. 
We plan to build on the original footprint of the Linear Collider Facility using technical innovation without requiring new civil construction. 
Decisions for these stages can be deferred until the 2030s and beyond.  
We will describe some of the upgrade possibilities in Sec.~\ref{sec:upgrades}.

\paragraph{The first-stage LCF configuration} We propose a facility with a site length of \SI{33.5}{km}, including a \SI{5}{km} beam delivery region sized for up to \SI{3}{\TeV} collisions.
We plan that the project will include two interaction points, which would share the collider luminosity. 
We note that both ILC and CLIC have produced beam delivery system (BDS) designs with two interaction points.\footnote{The ILC BDS has been designed for up to \SI{1}{\TeV}, the CLIC BDS for up to \SI{3}{\TeV}. The precise BDS design for our proposal would still need to be optimised based on the existing ILC/CLIC designs.}
For the initial technology, we propose superconducting cavities with a gradient of \SI{31.5}{MV/m}, as developed for the ILC, but with a quality factor $Q_0$ of $2\times10^{10}$. 
At its first stage, this facility would be equipped to reach a centre-of-mass energy of \SI{250}{\GeV}, with both beams polarised ($|P(\Pem,\Pep)|=(80\%,30\%)$), leaving part of the tunnel equipped only with a transfer line. As we will discuss below, this approach allows to flexibly increase the centre-of-mass energy up to \SI{550}{\GeV} by installing more accelerating modules at any time, at a speed adjustable to availability of resources (e.g.\ from non-member-states, which in this scenario could contribute additional cryomodules from their local industries in-kind) and scientific competition.

The RF and cooling systems will provide for trains of $1312$ bunches ($n_{e}=\num{2e10}$ per bunch) with a repetition rate of \SI{10}{Hz}, resulting in an instantaneous luminosity of \SI{2.7e34}{cm^{-2}s^{-1}}, a factor of two higher than for the ILC as proposed in Japan. 
The construction cost of this baseline configuration, without the detectors, is \SI{8.3}{BCHF}; the total required AC site power is \SI{143}{MW}. 

An attractive upgrade, either immediately if resources allow, or after a few years of running, is a further doubling of the luminosity to \SI{5.4e34}{cm^{-2}s^{-1}} by doubling the number of bunches per train. This would add \SI{0.77}{BCHF} to the accelerator construction cost, and increase the required site power to \SI{182}{MW}. 

On the other hand, reducing the facility length to \SI{20.5}{km}, imposing the use of a novel technology (c.f.\ Sec.~\ref{sec:upgrades}) to go beyond a centre-of-mass energy of \SI{250}{\GeV}, would reduce the initial cost by about \SI{0.7}{BCHF}; removing the second beam delivery system saves about \SI{0.3}{BCHF}; reducing the repition rate to \SI{5}{Hz} saves about \SI{0.5}{BCHF}. Thus a minimal \SI{250}{\GeV} machine would cost \SI{6.8}{BCHF}.

\paragraph{The second-stage LCF configuration}  In the \SI{33.5}{km} facility, a centre-of-mass energy of at least \SI{550}{\GeV} can be reached by equipping the additional tunnel length with further SCRF accelerator modules.  
This would require an additional \SI{5.5}{BCHF}, assuming that the same production facilities as for the initial stage cryomodules can be reused. 
With a total AC power of \SI{322}{MW} an instantaneous luminosity of $7.7\times 10^{34}$\,cm$^{-2}$s$^{-1}$ would be achieved. 
After several years of experience with operating the positron source in the more challenging \SI{250}{\GeV} configuration, the positron polarisation can be increased to $|P(\Pep)|=60\%$~\cite{Adolphsen:2013kya} for the \SI{550}{\GeV} stage. Table~\ref{tab:lcf-params} summarises some key parameters for these scenarios.  A more comprehensive version of the table can be found in the back-up document~\cite{LCF:EPPSU:backup}.

\paragraph{Run Plan} Figure~\ref{fig:runplan:base} shows a possible run plan starting with the \SI{250}{\GeV} machine with 1312 bunches per train, and including a luminosity upgrade to 2625 bunches per train as well as an energy upgrade to \SI{550}{\GeV}. Figure~\ref{fig:runplan:lup} shows the impact on the real time needed for the same total luminosities, but starting immediately with 2625 bunches per train.  
More possible data-taking scenarios are discussed in~\cite{LCVision-Generic}.

\begin{table}
\begin{tabular}{lcc|ccc|cc}
Quantity & Symbol & Unit & Initial-250 &  \multicolumn{2}{c|}{Upgrades} & Initial-550&  Upgrade\\
\hline
Centre-of-mass energy & $\sqrt{s}$ & ${\mathrm{\GeV}}$ & $250$ & $250$ & $550$ & $550$  & $550$ \\
Inst. Luminosity & \multicolumn{2}{c|}{${\mathcal{L}}$ $(10^{34}{\mathrm{cm^{-2}s^{-1}}}$)}& $2.7$ & $5.4$ & $7.7$ & $3.9$ & $7.7$ \\
Polarisation & \multicolumn{2}{c|}{$|P(e^-)|$/ $|P(e^+)|$ (\%)} & 80 / 30 & 80 / 30 & 80 / 60  & 80 / 30 &  80 / 60 \\ \hline 
Bunches per pulse  &$n_{\mathrm{bunch}}$ & 1  & $1312$ & $2625$ & $2625$ &  $1312$ & $2625$  \\
Average beam power  & $P_{\mathrm{ave}}$   & ${\mathrm{MW}}$ & $10.5$  & $21$ & $46$ & $23$ & $46$\\  
Site AC power  & $P_{\mathrm{site}}$ &  ${\mathrm{MW}}$ & $143$ & $182$ & $322$ & $250$ & $322$\\
Construction cost & & \si{BCHF} & 8.29 & +0.77 & +5.46 & 13.13 & +1.40 \\\hline
\multicolumn{2}{l}{Operation \& maintenance } & \unit{MCHF/y} & 170 & 196 & 342 & 291 & 342 \\
Electricity & & \si{MCHF/y} & 66 & 77 & 142 & 115 & 142 \\
Operating Personnel & & \unit{FTE} & 640 & 640 & 850 & 850 & 850 \\
\end{tabular}
\caption{Summary table of the LCF accelerator parameters in the initial \SI{250}{\GeV} configuration (4th column) and possible upgrades (first to 2625 bunches per train, then to \SI{550}{\GeV}), as well as a configuration for starting directly at \SI{550}{\GeV}   with 1312 bunches (7th column) and its upgrade to 2625 bunches per train. All options based on a \SI{33.5}{km} long facility.
\label{tab:lcf-params}}
\end{table}

\begin{figure}[htb]
\begin{center}
  \begin{subfigure}{.5\textwidth}
    \centering
    \includegraphics[width=0.90\hsize]{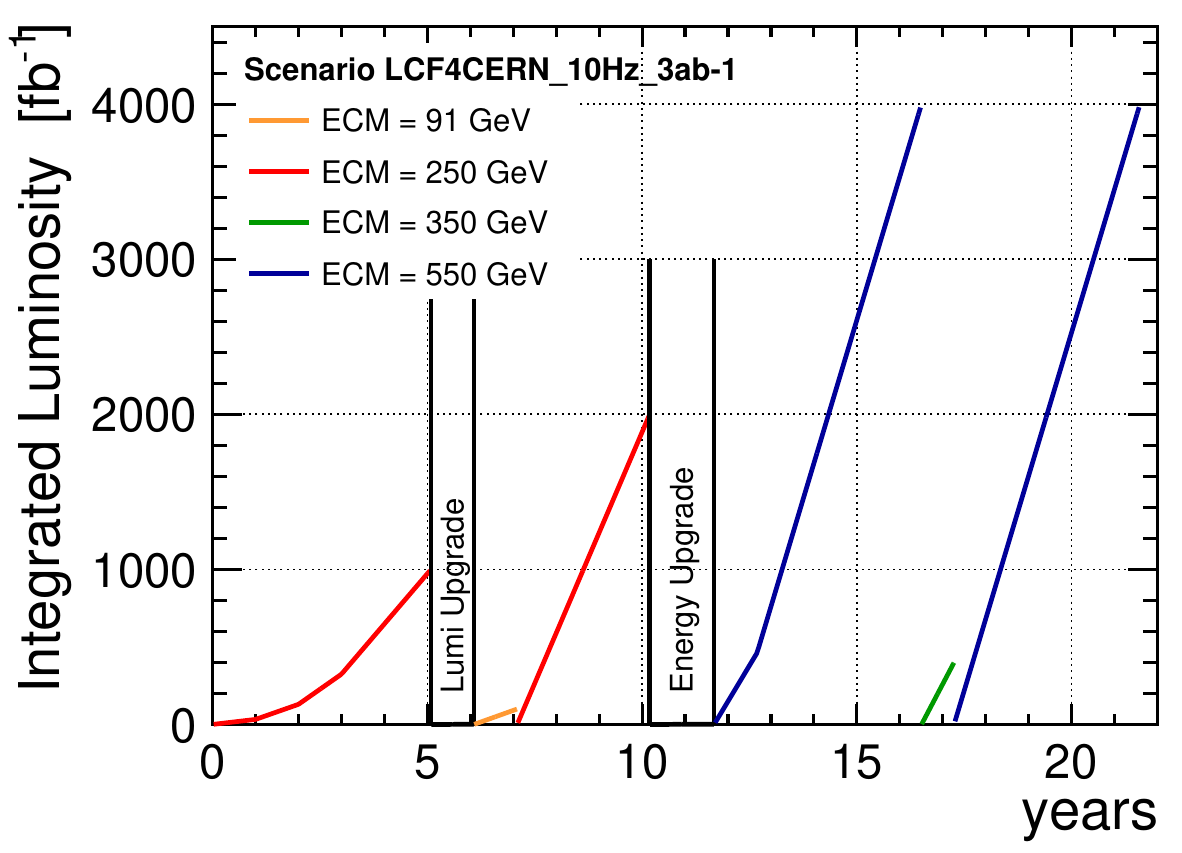}
    \caption{}
    \label{fig:runplan:base}    
  \end{subfigure}\hfill%
  \begin{subfigure}{.5\textwidth}
    \centering
    \includegraphics[width=0.90\hsize]{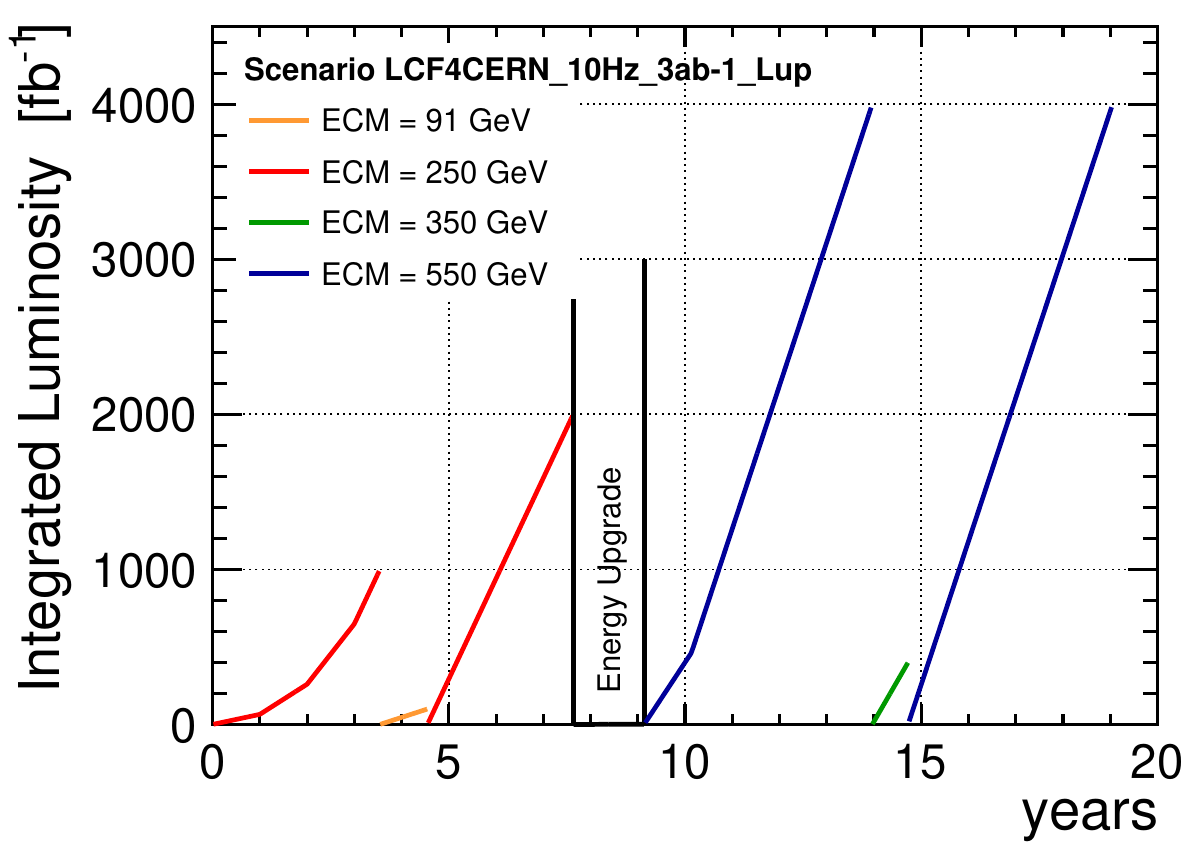}
    \caption{}
    \label{fig:runplan:lup}    
  \end{subfigure}\hfill%

\caption{\label{fig:runplan} Two of the suggested run plans for a CERN linear collider, based on $1.2\times10^{7}$\,s operation time per year, and including assumptions on performance ramp-up analogously to~\cite{Barklow:2015tja}. (a) starting in the ``low-power'' configuration (b) starting directly with the ``full-power'' option. The technical time zero would be close to 2044. 
}
\end{center}
\end{figure}

In Fig.~\ref{fig:projections}, we showed the expected uncertainties on Higgs boson couplings that the run plan in Fig.~\ref{fig:runplan} would provide. 
We highlight here that with the envisioned \SI{8}{\abinv} at \SI{550}{\GeV} and $|P(e^-,e^+)| = (80\%,30\%)$, the Higgs self-coupling can be measured with a precision of \SI{11}{\%}, and the top Yukawa coupling to \SI{2}{\%}, both improving further with $|P(e^+)| = 60\%$.

\paragraph{Start at 550\,GeV} Should the scientific and/or strategic need arise, the purchase and installation of a sufficient number of acceleration modules for reaching \SI{550}{\GeV} (or as an intermediate step the $\PQt\PAQt$ production threshold) could be advanced -- provided sufficient resources are made available. 
In such a scenario, the operation of the collider facility could start directly at \SI{550}{\GeV}, providing a set of measurements complementary to those at lower centre-of-mass energies, listed in the second bullet of Section~\ref{sec:phys}. 
This starting point offers the flexibility to optionally collect polarised data at lower energies if the scientific need arises, or to directly increase the centre-of-mass energy further with one of the technology upgrade options discussed in the following section, addressing the programme outlined in the third bullet of Section~\ref{sec:phys}. 
Constructing the \SI{550}{\GeV} machine immediately would require \SI{13.8}{BCHF} for operation with 1312 bunches per train, which includes costs for creating twice the production capacity. Upgrading the power to double the luminosity (2625 bunches per train) costs an additional \SI{1.4}{BCHF}.

\paragraph{Beyond a collider} A linear collider facility offers the opportunity to cater to a diverse "beyond collider" programme, including beam dump experiments with several $10^{21}$ electrons / positrons on-target per year, experiments with beam extracted before the collisions for fixed target experiments, test beams and irradiation facilities as well as large scale accelerator R\&D. 
The R\&D  and physics opportunities, spanning from light new particle searches to strong-field QED,  are described in more detail in~\cite{LCVision-Generic}. The necessary underground construction is best done together with the initial facility, based on a design to be developed together with the "beyond colliders" community. 
Since no detailed designs exist yet, the "beyond colliders" facilities are not included in the costing.


\section{Upgrade Scenarios for the Linear Collider Facility}
\label{sec:upgrades}
As we emphasised in the previous section, energy upgrades of the  Linear Collider Facility would be achieved with technological progress rather than with new civil construction. The proposed facility offers numerous options to increase the energy, the luminosity and/or the type of colliding particles. Choices can be made at later stages taking into account scientific and technological developments -- or even revolutions. We give here a brief summary of the opportunities described in~\cite{LCVision-Generic}.

\subsection{Energy upgrades beyond 550\,GeV}

There are four technical solutions now under consideration to reach much higher accelerating gradients, targeting gradients of at least \SI{70}{MV/m}, reaching up to \SI{155}{MV/m}. 
An accelerator with these gradients would reach a centre-of-mass energy between \SI{1} and \SI{3}{\TeV} in the \SI{33.5}{km} tunnel. 
For the first two technologies, this gradient has already been achieved and substantially exceeded in test modules.
All four technologies have fully engaged R\&D programmes that should reach maturity by the early 2030s. 
Once the construction of the first-stage LCF has begun, CERN can evaluate these options to find the best solution.  
One should remember that, in the ILC design, the actual linear accelerator accounts for less than half the project cost. 
Depending on the chosen technology, not only the tunnel but also damping rings, transfer lines, BDS elements etc., can be re-used. 
Such an upgrade is expected to be substantially more cost-effective and sustainable than building a facility from scratch. 
The LCVision document~\cite{LCVision-Generic}  gives a more detailed description of the accelerator technologies, status of the accelerator R\&D, and plans for re-use of the first-stage linear collider elements in each of these cases. 

\begin{itemize}

  \item { \bfseries CLIC technology}:   
    By replacing the SCRF linacs with an X-band accelerator with CLIC-like cavities~\cite{Brunner:2022usy}, running at a
    gradient of \SI{72}{MV/m} a centre-of-mass energy of about \SI{1.5}{TeV} could be reached in the LCF.

     \item { \bfseries C$^3$ technology}: Alternatively, the SCRF linacs could be replaced with a C-band copper accelerator using the innovations of distributed coupling and cryogenic liquid N$_2$ operation envisioned for the C$^3$ proposal~\cite{vernieri2023cool, Vernieri:2022fae}. 
     Depending on the future achievable gradient and advances in high-efficiency RF sources, centre-of-mass energies between \SI{1.5}{TeV} and \SI{3}{TeV} could be reached in the LCF~\cite{CCC:ESPPU}.
   \item {\bfseries HELEN technology}: By replacing the initial SCRF with travelling-wave Nb cavities with gradients of \SI{60}{MV/m}, like in the HELEN proposal~\cite{Belomestnykh:2023uon, Belomestnykh:2023naf}, a centre-of-mass energy of at least \SI{1}{TeV} could be reached in the LCF.
   \item {\bfseries Nb$_{\mathrm{\boldmath 3}}$Sn technology}: 
   Advanced superconducting technology based on the use of Nb$_3$Sn~\cite{PosenSnowmass2021} can reach gradients as high as \SI{90}{-} \SI{100}{MV/m}. 
   Once these become application-ready,  a centre-of-mass energy of at least \SI{1.5}{TeV} could be reached in the LCF.
\end{itemize}

\subsection{Additional upgrade paths}

Either further raising centre-of-mass energy or addressing other technical goals may benefit the study of Higgs bosons and other particle physics reactions.  
Solutions directed at these goals are less developed than the  options just discussed, but they could become relevant in later stages of the Higgs factory programme.

A linear collider with two interaction regions can share luminosity between these regions in an arbitrary ratio, and can also allow the two regions to see collisions of different kinds of particles, e.g.\ photons.
Alternatively to sharing the $\ee$ luminosity equally between two detectors, the CERN Linear Collider Facility could operate with one interaction region dedicated to accumulating luminosity while the other interaction region serves also as a development platform for new concepts, which eventually would be applied in upgrades of the facility.

There are three approaches that are now under consideration (for more details, and in particular a discussion of their application in the context of the CERN Linear Collider Facility, see~\cite{LCVision-Generic}):

\begin{itemize}

  \item {\bfseries  Photon collider}:  Observing the Higgs boson in
    $\PGg\PGg$ collisions adds a new set of processes with  powerful observables complementary to
     those available in $\ee$ collisions. In particular, the Higgs self-coupling can be 
     (re-)measured at lower   centre-of-mass 
     energy in $\PGg\PGg \to \PH\PH$. At the LCF, one of the interaction points could host a $\PGg\PGg$ collider either based on the classic scheme with optical lasers or on X-ray lasers as in the  XCC design~\cite{Barklow_2023}. 

\item {\bfseries Energy and particle recovery}:  At a linear collider, the luminosity is proportional to the beam power.  
A superconducting accelerator can in principle recover and reuse the beam  power and even the beam particles, to produce luminosities up to  \SI{e36}{cm^{-2}s^{-1}}, for instance in a scheme like ReLiC~\cite{Litvinenko:2022qbd} or ERLC~\cite{Telnov_2021}.   

\item {\bfseries Plasma wakefield acceleration}:  By offering accelerating gradients of several GV/m, this technology opens a route to much higher energies.  At the LCF, it could provide higher energies by adding plasma cells to the electron arm of a conventional accelerator producing asymmetric collisions \`a la HALHF~\cite{Foster:HALHF2023, Foster:HALHF2025, HALHF:EPPSU}. For the longer future, it may make possible $\ee$ or $\PGg\PGg$ colliders with energies of \SI{10}{TeV} and above~\cite{10TeV_AAC, Cros:2019tns, Cros:2019tns, ALEGRO:EPPSU}. At CERN, also p-driven PWA~\cite{ALIVE:EPPSU} is an option.

 \end{itemize} 

  \subsection{Beyond the Linear Collider Facility}
As ``high-energy'' physicists, we are dedicated to the exploration of increasingly higher energies.  It is important, then, to have a path to a collider operating at the \SI{10}{TeV} parton energy scale. 
Today, we do not have any accelerator technology that can achieve this for a reasonable cost.   
Substantial, and expensive, accelerator R\&D is required.   
Three paths are being pursued, with high-field magnets for a proton collider, with muon cooling for a muon collider, and with plasma wakefield acceleration for an $\ee$ or $\PGg\PGg$  collider.  
Funding for all three R\&D paths needs to be increased. 
It is important that CERN retain enough financial headroom beyond immediate projects to undertake this R\&D. 
This will be especially critical when these approaches require major technology demonstrators. 
In the 2030s, we will see a competition for resources between Higgs factory construction and R\&D toward a \SI{10}{\TeV} technology. 
We are pleased that R\&D for all three approaches is pursued in the framework of the accelerator roadmap of the European Lab Directors Group~\cite{Adolphsen:2022ibf}. 
The cost must be shared globally, but Europe can best play its part by opting for a staged and flexible Higgs factory plan.


\section{Next Steps Towards a Linear Collider Facility for CERN}
\label{sec:next}
The project implementation for the LCF foresees two preparation phases of three and five years, respectively: 
\vspace{-0.2cm}
\paragraph{Phase~1,} ideally starting directly in 2026 after the finalisation of the EPPSU,  
builds on and integrates
the on-going work of the ILC Technology Network (ITN) on the key technologies~\cite{IDT-EB-20023-002, ILC-EPPSU:2025}, complementing it with siting and implementation studies, in parallel with design and technical studies to determine and confirm the final LCF parameters. This phase is required to prepare a project decision by CERN Council, which is not expected before 2028, and should deliver the following:
\begin{itemize}

\item {\bfseries Final placement of the collider complex} and its infrastructure after assessment of their territorial compatibility, both for its initial phase and potential upgrades.  Preparation and documentation for wider implementation studies with the host regions / states in Phase~2. 

\item {\bfseries Optimisation of LCF design:} Review of the accelerator design along with updated cost, power and risk assessments, including in particular
\begin{itemize}
    \item a detailed study of the interaction region for two experiments -- including re-optimised beam delivery systems -- that is suitable for an initial SCRF-based accelerator as well as for future technology upgrades like CLIC, C$^3$, ERLs or PWA; 
    \item studies of the implications for civil engineering and the equipment used in the initial SCRF-based facility, in order to allow collider upgrades in the future, and to accommodate the "beyond colliders" physics described in Sec.~\ref{sec:upgrades};
    \item design and parameter optimisation of the entire machine including further nanobeam studies, and further R\&D on key components.
\end{itemize}

\end{itemize}

The estimated resources needed for Phase~1 are \SI{35}{MCHF} material and \SI{180}{FTEy} of personnel effort, in addition to the ITN efforts that are of the same size.

\paragraph{Phase~2} requires a formal decision to proceed with the project. It targets the final engineering design, larger industrial pre-series and an extensive site preparation, in particular

\begin{itemize}
    \item SCRF industrialisation of a design based on more ambitious goals for cavity quality factors ($Q_0$) and power (klystron) efficiency, with higher repetition rate and hence cryogenic needs than foreseen in the ILC TDR design. This task builds on the work of the ITN~\cite{IDT-EB-20023-002};
    \item detailed studies of the cryogenics system and infrastructure systems (e.g.\ cooling and ventilation, electrical, access and safety systems, transport and installation) adapting them to standard solutions used by CERN and industry;
    \item final site preparation including documentation and specifications for the time-critical civil engineering contracts. 
    \item environmental studies and integration of the collider in the local area with the host states.
\end{itemize}

The estimated resources needed for Phase~2 are \SI{120}{MCHF} for pre-series production and \SI{420}{FTEy} of personnel effort for the technical studies, pre-series, engineering design and laboratory infrastructure. With the widespread expertise built up worldwide for SCRF-based free-electron lasers (e.g.\ the Eu.XFEL or LCLS-II) as well as for the ILC, a significant part of this work can be done outside CERN for an LCF starting with SCRF technology. In parallel, civil engineering preparation including continued environmental studies will require significant resources, typically 5\% of the civil engineering budget.

During the preparation period, detector collaborations need to be prepared and set up, building on 
the well developed detector concepts for Higgs factories, yet embracing new ideas.

Construction could start earliest in 2034, preceded by a formal decision by CERN Council. Note further that in the current planning laid out in the addendum, the construction time is increased beyond the purely technical limitation of eight years to ten years in order to accommodate the fact that the transition between phases might require some time, and also to avoid the clear conflict between HL-LHC operation and beam commissioning of a new collider.


\section{Conclusions}
\label{sec:conclusions}
In this report, we have outlined a proposal for a cost-effective, staged construction of a Linear Collider Facility at CERN that will be able to comprehensively map the Higgs boson's properties, including the Higgs field potential, and to account for potential discoveries at the HL-LHC.
The need for a Higgs factory for the global particle physics community  is urgent.  
This is the key to solving the most important mysteries left unresolved by the Standard Model.   
The Higgs boson is the most likely particle to couple to new fundamental interactions, yet, still, it is the least studied. 

The programme that we have put forward, spanning centre-of-mass energies from the $\PZ$ pole to at least \SI{1}{\TeV}, is attentive to the requirements for making a true discovery from precision measurements.  
Its variety of precision observables allows deviations from the Standard Model to be measured in different ways that can be cross-checked against one another. 
The discovery of deviations from the Standard Model -- or, better, a pattern of deviations in different Higgs couplings -- will guide us to the discovery of new fundamental laws operating at higher energies.

A new generation of particle physicists, brought up with the LHC, will need a well-defined and timely plan towards new opportunities once the HL-LHC ends. 
Our staged and flexible programme provides a conservative yet upgradable starting point that can be constructed on the timescale of the HL-LHC, and can be the site to realise ambitious new accelerator and detector technologies.  
It can host continued strong R\&D on advanced accelerator concepts and a suite of smaller-scale experiments.  
The advanced stages of our programme will give targets for technical developments  that will  be useful to the particle physics community, to other scientific fields, and potentially to society in general.

The Linear Collider Facility proposed in this document will be an excellent flagship project for CERN.  
It addresses all of the issues associated with the next CERN project and provides a route to the higher-energy colliders of the future. 
This is the project that CERN and the global particle physics community needs now. 

\section*{Acknowledgements}

{\bfseries
The LCVision Team acknowledges with deep gratitude that putting together this LCF proposal in less than one year would not have been possible without the decades of excellent scientific and technical work of the whole Linear Collider community, and in particular that of the ILC Global Design Effort, the Linear Collider Collaboration,  the ILC International Design Team, and the CLIC Collaboration.}\\

This work was supported by EAJADE, a Marie Sklodowska-Curie Research and Innovation Staff Exchange (SE) action, funded by the EU under Horizon-Europe Grant agreement ID: 101086276; by AIDAinnova, a project within the European Union's Horizon 2020 Research and Innovation programme under GA no. 101004761; 
by the CNRS/IN2P3, France; 
by the Deutsche Forschungsgemeinschaft (DFG, German Research Foundation) under grant 491245950 and under Germany's Excellence Strategy -- EXC 2121 ``Quantum Universe'' -- 390833306, 
and the DFG Emmy Noether Grant no.\ BR 6995/1-1; 
by a Department of Science and Technology and Anusandhan National Research Foundation Government of India Startup Research Grant, grant agreement no. SRG/2022/000363 and Core Research Grant, grant agreement no. CRG/2022/004120; 
by the National Science Centre (Poland) under the OPUS research project no. 2021/43/B/ST2/01778  
and by Narodowe Centrum Nauki, Poland, grant no. 2023/50/A/ST2/00224; 
by the Spanish Ministry of Science under grant agreement PID2021-122134NB-C21, by the Generalitat Valenciana under CIPROM/2021/073 and ASFAE2022/013 and 015, by the Severo Ochao excellence program,  
and by the Atracci\'on de Talento Grant no. 2022-T1/TIC-24176 of the Comunidad Aut\'onoma de Madrid, Spain; 
by the Swiss National Science Foundation under grant no. 214492; 
by the Science and Technology Facilities Council, United Kingdom; 
by the US Department of Energy under contracts No. DE-AC02-76SF00515,  no. 89243024CSC000002,  
no. DE-SC0010107, and
no.~DE-SC0010107,  
by the US National Science Foundation through the award NSF2310030, and by the Los Alamos National Laboratory LDRD programme.

\section*{Appendices}
{\appendix
\section{Stages and parameters}

\paragraph{The main stages of the Linear Collider Project (LCF) and the key scientific goals of each} are sketched in Fig.~\ref{fig:graphscenarios} for two scenarios A and B, which are defined as follows: 

\begin{figure}[tbh]
    \centering
    \includegraphics[width=0.90\hsize]{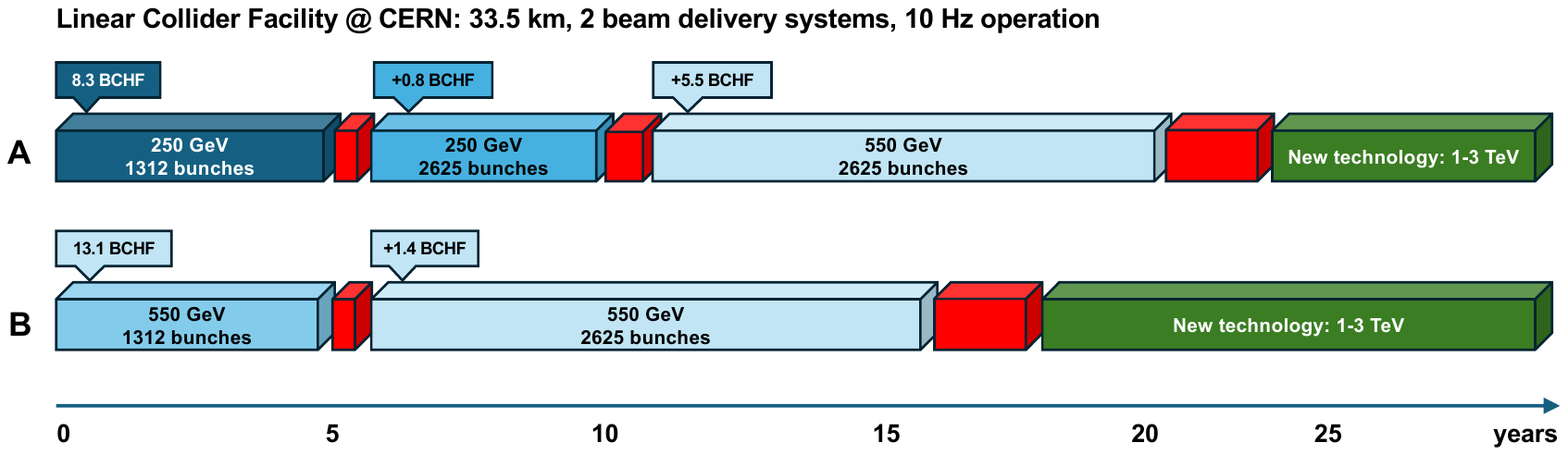}
\caption{\label{fig:graphscenarios} Visualisation of the timelines of scenarios A and B. 
The technical time zero would be around 2044. 
Note that the Linear Collider Facility offers significant flexibility to react to i) findings at or developments with other machines (HL-LHC, CEPC, ...) and to ii) the technical readiness of particular implementations.  }
\end{figure}

\begin{itemize}
\item \textbf{Scenario A}, i.e.\ starting with \SI{250}{GeV}, gives a large data set of $\ee\rightarrow \PZ\PH$ events for measurements of Higgs branching ratios (to \SI{1}{\%}), the total $\PZ\PH$ cross section (to \SI{1}{\%}), the absolute Higgs coupling normalisations and the Higgs boson mass (to $10^{-4}$ precision). It also enables  searches for non-standard Higgs boson decays (to the $10^{-4}$ level of branching ratios), and it allows for a collection of a few $10^9$ events at the $\PZ$ pole. 

\begin{itemize}
\item The \textbf{power upgrade of scenario A} from 1312 ("low power", LP) to 2625 bunches ("full power", FP) will double the available luminosity. 

\item The \textbf{energy upgrade of scenario A} to \SI{550}{GeV} will give access to $\PW\PW$ fusion events, with an independent setting for Higgs branching ratio measurements and exotic decay searches. 
Importantly, it will facilitate measurements of the top quark Yukawa coupling (through $\ee\rightarrow \PQt\PAQt\PH$) and of the Higgs self-coupling (through $\ee\rightarrow \PZ\PH\PH$). 
It will also enable measurements of top quark properties, not least the top mass in top-pair threshold measurements. 

\end{itemize}

\item \textbf{Scenario B}, i.e.\ starting with \SI{550}{GeV}, gives immediate access to all mentioned measurements in the entire energy range from the $\PZ$ pole to \SI{550}{GeV}.
\begin{itemize}

\item The \textbf{power upgrade of scenario B}, from 1312 to 2625 bunches, will double the available luminosity. 

\end{itemize}

\item A later \textbf{stage at \SI{1}{TeV}} or beyond provides access to $\PW\PW$ fusion to $\PH\PH$ and thus a complementary access to the Higgs self-coupling. 
There will be increased sensitivity to anomalies in the top quark couplings. 
Ample single-Higgs production will allow for further scrutiny of the Higgs couplings to SM particles. 

\item An potential ultimate \textbf{stage at \SI{3}{TeV}} might grant access to new particles discovered or hinted to at the HL-LHC, and will give access to the quartic Higgs coupling.

\end{itemize}

\begin{table}
\centering
\footnotesize
\begin{tabular}{lcc|ccc|cc}
Quantity & Symbol & Unit & Initial-250 &  \multicolumn{2}{c|}{Upgrades} & Initial-550 &  Upgrade\\
Name &  & LCF  & 250 LP & 250 FP & 550 FP & 550 LP &  550 FP \\
\hline
Centre-of-mass energy & $\sqrt{s}$ & GeV & $250$ & $250$ & $550$ & $550$  & $550$ \\
Inst. luminosity & \multicolumn{2}{c|}{${\mathcal{L}}$ $(10^{34} cm^{-2}s^{-1}$)}& $2.7$ & $5.4$ & $7.7$ & $3.9$ & $7.7$ \\
Polarisation & \multicolumn{2}{c|}{$|P(e^-)|$/ $|P(e^+)|$ (\%)} & 80 / 30 & 80 / 30 & 80 / 60  & 80 / 30 &  80 / 60 \\ \hline 
Repetition frequency &$f_{\mathrm{rep}}$ & Hz  & $10$ & $10$ & $10$ & $10$  & $10$\\
Bunches per pulse  &$n_{\mathrm{bunch}}$ & 1  & $1312$ & $2625$ & $2625$ &  $1312$ & $2625$  \\
Bunch population  &$N_{\mathrm{e}}$ & $10^{10}$ &$2$ & $2$ & $2$ & $2$ & $2$\\
Linac bunch interval & $\Delta t_{\mathrm{b}}$ & ns & $554$ & $366$ & $366$& $554$  & $366$ \\
Beam current in pulse & $I_{\mathrm{pulse}}$ & mA & $5.8$ & $8.8$ & $8.8$ &  $5.8$ & $8.8$ \\
Beam pulse duration  & $t_{\mathrm{pulse}}$ & $\mu$s & $727$ & $897$ & $897$ & $727$ & $897$  \\
Average beam power  & $P_{\mathrm{ave}}$   & MW & $10.5$  & $21$ & $46$ & $23$ & $46$\\  
Norm. hor. emitt. at IP & $\gamma\epsilon_{\mathrm{x}}$ & $\mu$m & $5$ & $5$ & $10$ & $10$ & $10$ \\ 
Norm. vert. emitt. at IP & $\gamma\epsilon_{\mathrm{y}}$ & nm & $35$ & $35$ & $35$ & $35$ & $35$ \\ 
RMS hor. beam size at IP  & $\sigma^*_{\mathrm{x}}$ & nm  & $516$ & $516$ & $452$ & $452$ & $452$\\
RMS vert. beam size at IP &$\sigma^*_{\mathrm{y}}$ & nm & $7.7$ & $7.7$ & $5.6$ & $5.6$ & $5.6$ \\
Lumi frac. in top 1\,\% & ${\mathcal{L}}_{\mathrm{0.01}}/{\mathcal{L}}$  & \% & 73 & 73 & 58 & 58 & 58 \\
Lumi in top 1\,\% & \multicolumn{2}{c|}{${\mathcal{L}}_{\mathrm{0.01}}$ $(10^{34} cm^{-2}s^{-1}$)}& 2.0 & 4.0 & 4.5 & 2.2 & 4.5 \\
\hline
Site AC power  & $P_{\mathrm{site}}$ &  MW & $143$ & $182$ & $322$ & $250$ & $322$\\
Annual energy consumption &  & TWh & $0.8$ & $1.0$ & $1.8$ & $1.4$ & $1.8$ \\
Site length & $L_{\mathrm{site}}$ &  km & $33.5$ & $33.5$ & $33.5$ & $33.5$ & $33.5$\\
Average gradient & $g$ &  MV/m & $31.5$ & $31.5$ & $31.5$ & $31.5$ & $31.5$\\
Quality factor & $Q_0$ &  $10^{10}$ & $2$ & $2$ & $2$ & $2$ & $2$ \\
\hline
Construction cost & & BCHF & 8.29 & +0.77 & +5.46 & 13.13 & +1.40 \\
Construction labour & & kFTE y & 10.12 &  & +3.65 & 13.77 &  \\
\hline
Operation and maintenance & & MCHF/y & 156 & 182 & 322 & 273 & 322 \\
Electricity & & MCHF/y & 66 & 77 & 142 & 115 & 142 \\
Operating personnel & & FTE & 640 & 640 & 850 & 850 & 850 \\
\end{tabular}
\caption{Summary table of the LCF accelerator parameters in the initial \SI{250}{GeV} configuration and possible upgrades, as well as in an initial \SI{550}{GeV} configuration and its luminosity upgrade.
\label{tab:lcf-params}}
\vspace{-0.5cm}
\end{table}

\begin{figure}[hhh]
    \centering
    \includegraphics[width=0.50\hsize]{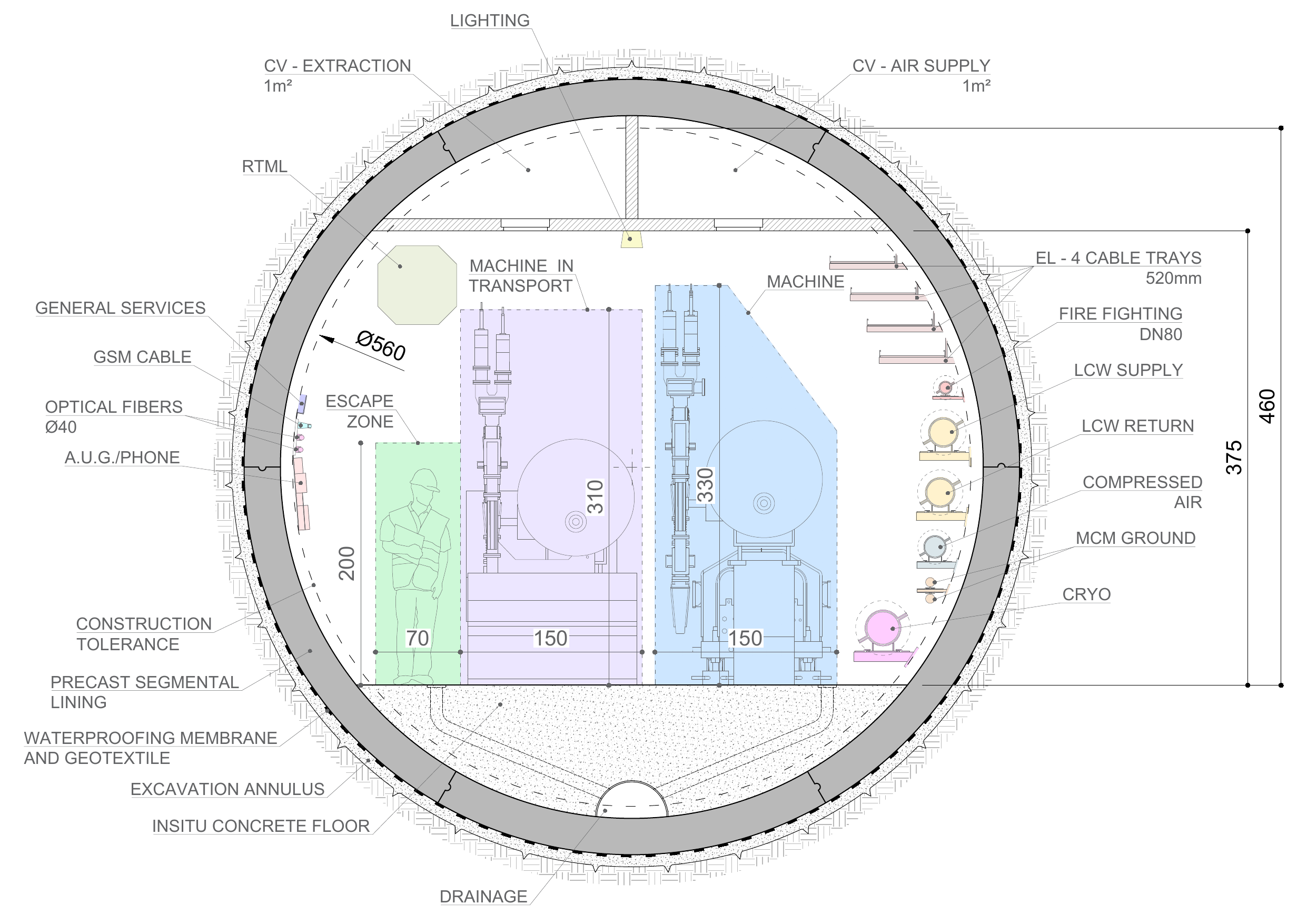}
\caption{\label{fig:tunnelxsection} Cross section of the \SI{5.6}{m} tunnel of a Linear Collider Facility at CERN.}
\end{figure}

\paragraph{The ordering and scope of stages is not fixed and allows for flexibility.} Two alternative starting scenarios for the LCF, one with \SI{250}{GeV} and one with \SI{550}{GeV} centre-of-mass energy, are sketched above. These scenarios are interchangeable, depending e.g.\ on the initially available funding. There is no technical reason preventing alternative starting scenarios, for instance directly with full power (i.e.\ twice higher luminosity) or at an intermediate centre-of-mass energy, e.g.\ \SI{380}{GeV}, provided sufficient initial funding can be secured. Note that thus the Linear Collider Facility in general offers significant flexibility to react to i) findings at or developments with other machines (HL-LHC, CEPC, ...) and to ii) the technical readiness of particular implementations.

\begin{figure}[hhh]
    \centering
    \includegraphics[width=0.95\hsize]{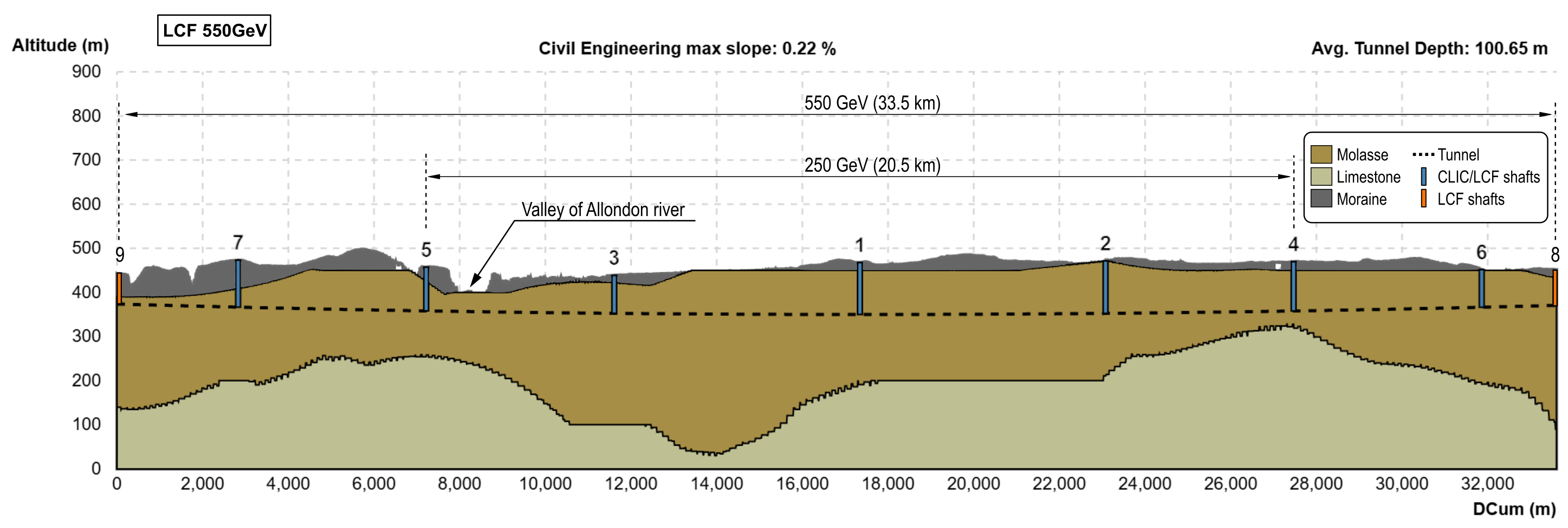}
\caption{\label{fig:geoprofile}. Geological profile of the LCF tunnel and the location of its access shafts.}
\end{figure}

\paragraph{For each stage, in both scenarios, the main technical parameters} are listed in Tab~\ref{tab:lcf-params}. 
As an important ingredient, Fig.~\ref{fig:tunnelxsection} shows the tunnel cross-section of diameter \SI{5.6}{m} for the Linear Collider Facility at CERN. Figure~\ref{fig:geoprofile} shows the geological profile of the foreseen LCF tunnel location, together with the indicative shaft locations. 


\section{\label{sec:timeline}Timeline}

\textbf{Both scenarios A and B are essentially ready to be built} (see the discussion of remaining R\&D items in Sec.~\ref{sec:techdev}). 
The chosen technology -- ILC-type superconducing RF -- is mature and has its industrialisation has been successfully completed in the context of the European XFEL at DESY, Hamburg.  
A construction and site preparation time of 9 years is assumed, irrespective of the scenario. 
Table~\ref{tab:timeline_ILC} shows the anticipated timeline for the various steps of the project in years. 
The definition of project implementation phases is used as given in the main document of this submission.  T$_0$ is determined by a process in 2028-29 to validate the progress and promise of the project for a further development towards implementation. T$_1$ for the start of the construction phase will be determined by the processes needed, by the CERN Council and with host-states, for project approval and to start construction. The construction phase is extended with respect to the technical schedules to allow a transfer time into construction, and to avoid the resource conflict between HL-LHC operation and initiating beam commissioning for a next collider.

\begin{table}[h!]
\footnotesize
\centering
\begin{tabular}{|l|c|}\hline
Milestone & LCF@CERN \\ \hline\hline
Conceptual/Reference Design Report & 2002 -- 2007\\
Technical Design Report & 2007 -- 2013 \\ 
ILC 250 GeV reports and Prelab planning & 2013 -- 2025 \\

\hline
\hline
Project Preparation Phase 1 & 2026 -- 2028 \\
\hline
\multicolumn{2}{|c|}{Definition of the placement scenario}  \\

\multicolumn{2}{|c|}{Design optimisation and finalization}  \\

\multicolumn{2}{|c|}{Main technologies R\&D conclusions}  \\

\multicolumn{2}{|c|}{Technical Design Report -- two IPs at CERN} \\ 



\hline
\hline
Project Preparation Phase 2 & T$_0$ -- (T$_0$+5) \\
\hline 

\multicolumn{2}{|c|}{Site investigation and preparation}  \\
\multicolumn{2}{|c|}{Implementation studies with the Host states}  \\
\multicolumn{2}{|c|}{Environmental evaluation \& project authorisation processes} \\
\multicolumn{2}{|c|}{Industrialization of key components}  \\
\multicolumn{2}{|c|}{Engineering design completion}  \\

\hline
\hline
Construction Phase (from ground breaking) & T$_1$ -- (T$_1$+10) \\
\hline
\multicolumn{2}{|c|}{Civil engineering}  \\
\multicolumn{2}{|c|}{Construction of components}  \\
\multicolumn{2}{|c|}{Installation and hardware commissioning}  \\
\hline
\hline
Beam commissioning and physics operation start & T$_1+$11 \\
\hline
\end{tabular}
 \caption{Timeline of essential development and construction steps of the LCF project. See the text for details. 
 }
\label{tab:timeline_ILC}
\end{table}

\paragraph{The anticipated running times and envisaged integrated luminosities} for scenarios A and B are indicated in Fig.~\ref{fig:runplans}.
Note that these are tentative, and numerous other combinations of running periods are possible, depending on the physics landscape at the time. 
Not shown are \textbf{technology upgrades} to higher energies of \SI{1}{TeV} and beyond. 
For these, a construction and installation phase of 3 years each is foreseen. The necessary R\&D can be carried out in the shadow of running in scenario A or B. 

%
%
%
%
%
%
%

\begin{figure}[htb]
  \begin{subfigure}{.5\textwidth}
    \centering
    \includegraphics[width=0.90\hsize]{figures/lumi_LCF4CERN_10Hz_3ab-1.pdf}
    \caption{}
    \label{fig:runplan:base}    
  \end{subfigure}\hfill%
  \begin{subfigure}{.5\textwidth}
    \centering
    \includegraphics[width=0.90\hsize]{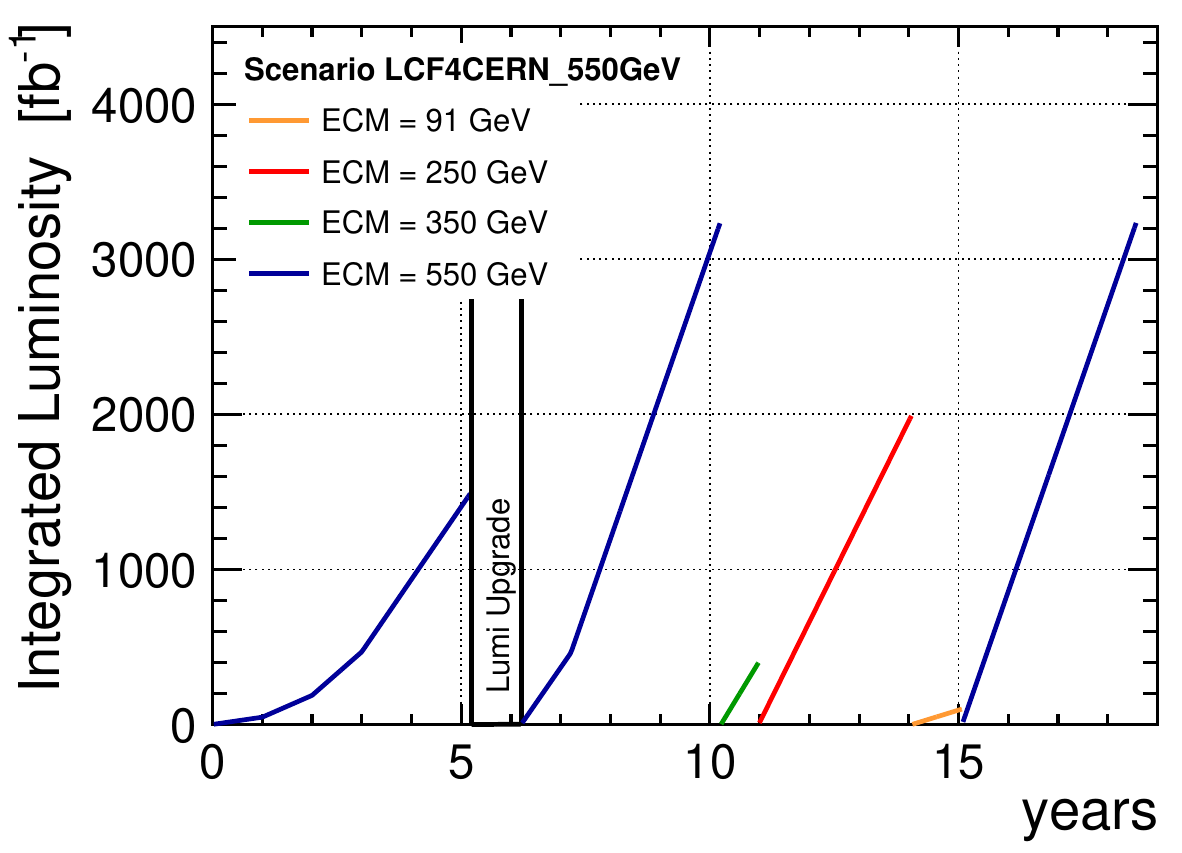}
    \caption{}
    \label{fig:runplan:lup}    
  \end{subfigure}\hfill%

\caption{\label{fig:runplans} 
Suggested run plans for an LCF at CERN (a) starting in the baseline configuration in scenario A, and (b) starting directly at \SI{550}{GeV}. 
The technical time zero would be around 2044.}
\end{figure}

\section{Resource requirements}
\subsection{\label{sec:methodology}Construction cost and costing methodology}

\paragraph{The capital cost} of all stages in the scenarios A and B sketched above as well as the annual costs of operation are given in Table~\ref{tab:lcf-params}. 
No cost estimates are given here for later stages of the project for energies beyond \SI{550}{GeV}.  

The human resources requirements for the construction of the initial stage of scenario A amount to \SI{10120}{FTEy}, and to around \SI{13770}{FTEy} for scenario B (i.e.\ when starting at \SI{550}{GeV}).

\paragraph{All cost estimates} given here for scenario A or B stages are based on the ILC technical design report (TDR)~\cite{Adolphsen:2013kya} from 2012 and several cost updates performed since then:  
\begin{itemize}
\item the cost estimate for a staged \SI{20.5}{km}, \SI{250}{GeV} configuration of the ILC as a Higgs factory in Japan from 2017~\cite{Evans:2017rvt}, which was based on the cost basis of the TDR estimate from 2012, but included the effect of a reduced length and centre-of-mass energy as well as a number of minor design updates;
\item the 2024 updated cost estimate of the ILC in Japan performed by the International Development Team (IDT)~\cite{ILC-EPPSU:2025,Yamamoto:cost-update};
\item a new cost evaluation of the construction costs for the CERN site from 2025~\cite{CERN-CFS-Cost-Update}.
\end{itemize}

\paragraph{The ILC cost estimate for the TDR}~\cite{Adolphsen:2013kya} is a full bottom-up cost estimate based on around 2000 line items, going down to the level of the cost of individual components such as cavities, couplers, or magnets for key cost drivers. 
The costs were evaluated and aggregated by an international team of experts and reviewed by an independent international expert panel in 2013.
The updated 2024 cost estimate was again reviewed by an international expert panel in Dec.\ 2024. A more detailed description of methodology, results and review are now contained in the addendum to the ILC submission to the EPPSU~\cite{ILC-EPPSU:2025}.

\paragraph{The scope of the cost estimate} encompasses the construction cost for the accelerator systems as detailed in the TDR, with applicable design changes.
Detectors are excluded, but detector assembly buildings, underground experimental halls, and detector access shafts are included.
Computing equipment for accelerator operation is included, while computing installations for detector operation, data taking and analysis are excluded.
The cost estimate covers the 10 year construction time until start of commissioning, and therefore excludes
costs for project engineering and design,  and for R\&D prior to construction authorization, as well as costs for 
commissioning, pre-operation, operation,  and de-commissioning.
Taxes, contingency, and escalation during project construction are not included.
Costs for upgrading the machine beyond \SI{550}{GeV} are not included, except the cost of those systems that would be very difficult to provide after construction of the \SI{550}{GeV} machine, in particular the main beam dumps and the beam delivery system (BDS).
Costs for land acquisition and site activation (external roads, water supplies, power lines) are not included. 

\paragraph{The scope of the new civil engineering cost estimate}~\cite{CERN-CFS-Cost-Update} encompasses the construction costs for underground buildings (tunnels, caverns, shafts) based on a design for the CERN site (Figs.~\ref{fig:tunnelxsection} and \ref{fig:geoprofile}), with tunnels and a central cavern for two beam delivery systems, adapted from the ILC TDR design to accommodate a possible later installation of CLIC. 
It also encompasses costs for accelerator-related surface buildings for cryogenics, cooling and power infrastructure, and for surface infrastructure in support of the detectors.

As discussed above, costs for land acquisition and site activation (external roads, water supplies, power lines) are \textbf{not included}, nor are costs for spoil removal.

The estimated amount of \textbf{spoil}, including a bulk factor of 1.3, is \SI{2.7e6}{m^3} (\SI{1.9e6}{m^3}) for the \SI{33.5}{km} (\SI{20.5}{km}) long facility.
The corresponding costs are estimated to be 200 and 140\,MCHF, respectively.

\paragraph{The methodology} adopted for the ILC cost estimate was chosen as deemed appropriate for the ILC as an international, in-kind contribution project, with no specific site or host country identified at the time of the TDR.
Civil infrastructure (separated in  costs for construction and for infrastructure such as electric power supply, water systems, heating, ventilation and air conditioning (HVAC) and several other systems) was evaluated for three potential sites, one in Japan, one in the United States, and one at CERN.
The costs were separated into ``Labour'' and ``Value'', where ``Labour'' corresponds to the total work performed by staff of participating institutions, measured in full-time equivalent person years (FTEy), while ``Value'' corresponds to the monetary value of goods and services procured from commercial vendors, expressed in ILCU2012 (ILC Currency Unit).
An ILCU2012 corresponds to the purchasing power of \SI{1}{USD} in the United States in Jan 2012.
Comparisons between prices obtained in different countries were made on the basis of PPP (purchasing power parity) conversion rates, as evaluated and publicised by the OECD and World Bank. 
Where price information from several vendors was available, in particular for SCRF components, prices were converted to ILCU2012 and the lowest reasonable price (where ``reasonable'' takes into account whether the offering company has been demonstrated the capability to meet the required specifications) was adopted as price estimate.
The information of the country of origin and currency of the original offer was retained, so that all quotations can be traced back to the original prices.

The \textbf{estimate for institutional Labour} of the \SI{550}{GeV} configuration was taken from the TDR estimate~\cite{Adolphsen:2013kya} and amounts to \SI{13.77}{kFTEy}, the estimate for the \SI{250}{GeV}  configuration is taken from the 2017 staging report~\cite{Evans:2017rvt} and amounts to \SI{10.12}{kFTEy}.

\paragraph{For the 2024 IDT cost update}~\cite{ILC-EPPSU:2025}, approximately 75\,\% of the costs were re-evaluated based on new quotations for SCRF components and updated civil engineering prices for the Japanese site.
The costs that were not re-evaluated covered the more conventional accelerator systems such as magnets, vacuum system, diagnostics, etc. 
Those costs were converted back to the original currency, escalated with the respective region's inflation index for machinery and equipment, and converted to an updated ILCU2024, defined as purchasing power of a USD in January 2024 in the United States. 

\paragraph{A cost estimate for the Linear Collider Facility at CERN} was derived based on the described inputs, with a number of important differences:
\begin{itemize}
    \item The Value estimate was converted to a procurement cost, expressed in 2024 Swiss Francs (CHF).
    \item The cost estimate was adjusted to take into account differences in the design of the facilities, such as different beam energies, the presence of a second BDS, and different RF configurations.
    \item The cost estimate for the construction costs of tunnels, caverns, shafts and surface buildings was based on a new study for the CERN site performed in early 2025.
    \item The cost estimate for the other civil engineering infrastructure (electric power, water, HVAC, etc.) was based on the costs evaluated in the 2012 TDR for the CERN site, escalated with the Swiss inflation index for machinery and equipment.
\end{itemize}

\paragraph{The conversion from Value to actual Cost in Swiss Francs} takes into account a procurement model appropriate for a CERN-based project, different from the ILC approach to quantify the Value of goods: 
The ILC is a project based on international in-kind contributions, where the procurement is done separately by the contributing institutions, subject to price differences and possibly with a motivation to procure locally at potentially higher prices.
This is a reason to consider the Value of contributed goods in lieu of the actual procurement cost in order to determine the monetary value of contributions to the overall project as laid out in a cost-book, which is not the same as the sum of all costs paid.
In contrast, a CERN-based project would predominantly organise a central procurement on the world market, and consider the cost of purchased goods, converted to CHF based on market currency exchange rates.

To convert the ILC Value numbers to \textbf{Cost expressed in CHF}, all item values were converted from ILCU2024 back to the local currency (i.e.\ US Dollars, Japanese Yen, Euros, or Swiss Francs) and then converted to CHF with exchange rates as of January 2024.
For goods that are available from different vendors around the world this is a conservative estimate insofar as the lowest reasonable price assumed in the ILC cost estimate is based on a PPP comparison, which leaves the possibility that the goods may be acquired in a different region of the world where price levels are lower than estimated from PPP rates. \\
\noindent
\textbf{The total cost for the baseline configuration LCF 250 LP amounts to \SI{8.29}{\textbf{BCHF}}, for the full configuration LCF 550 FP to \SI{14.53}{\textbf{BCHF}} in 2024 prices.}
These values are also summarised in Table~\ref{tab:lcf-params}, and they are broken down according to accelerator areas and to technical systems on the left and right side, respectively, of Fig.~\ref{fig:costs}.

\begin{figure}[htb]
    \centering
    \includegraphics[width=0.48\hsize]{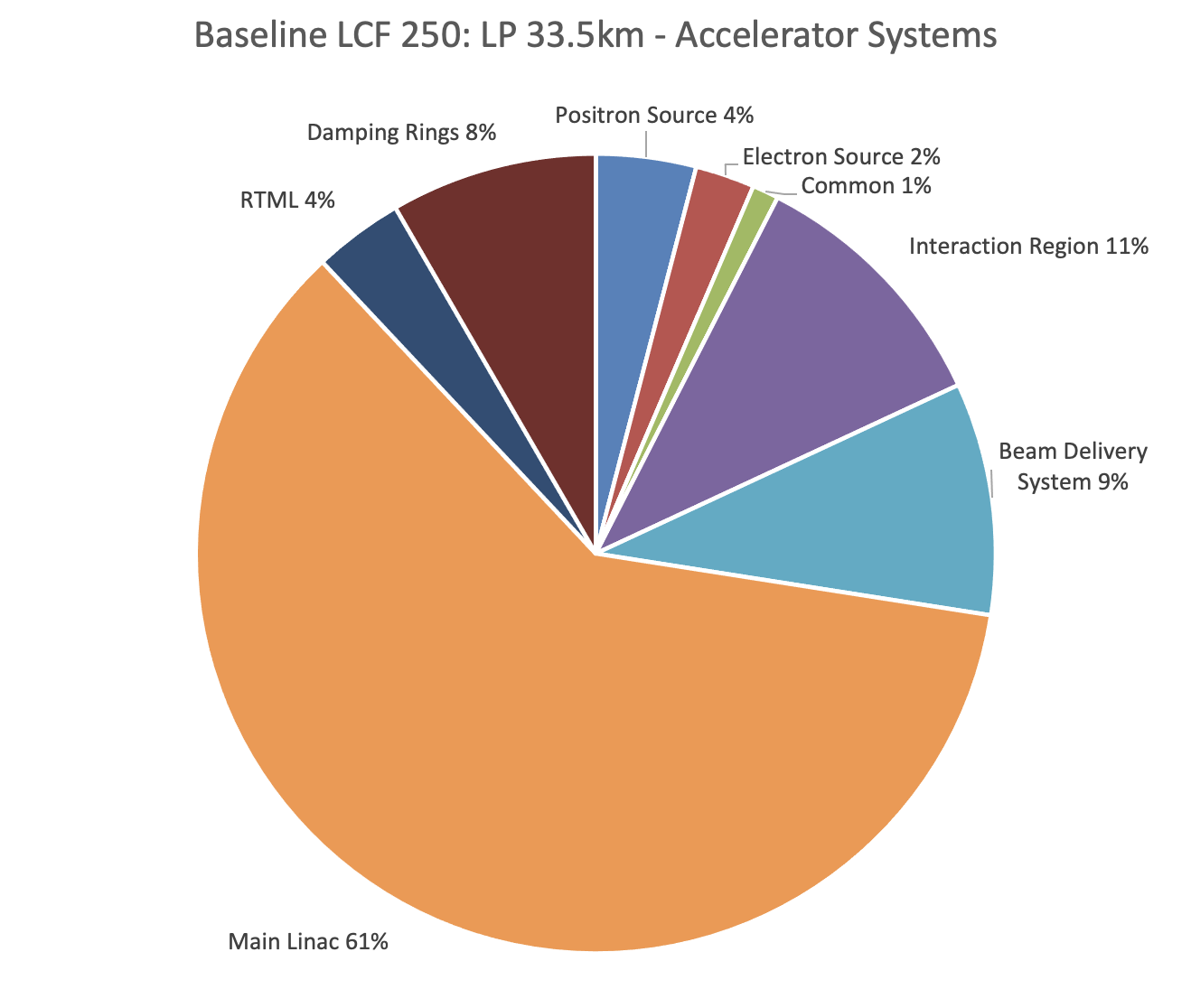}
    \includegraphics[width=0.48\hsize]{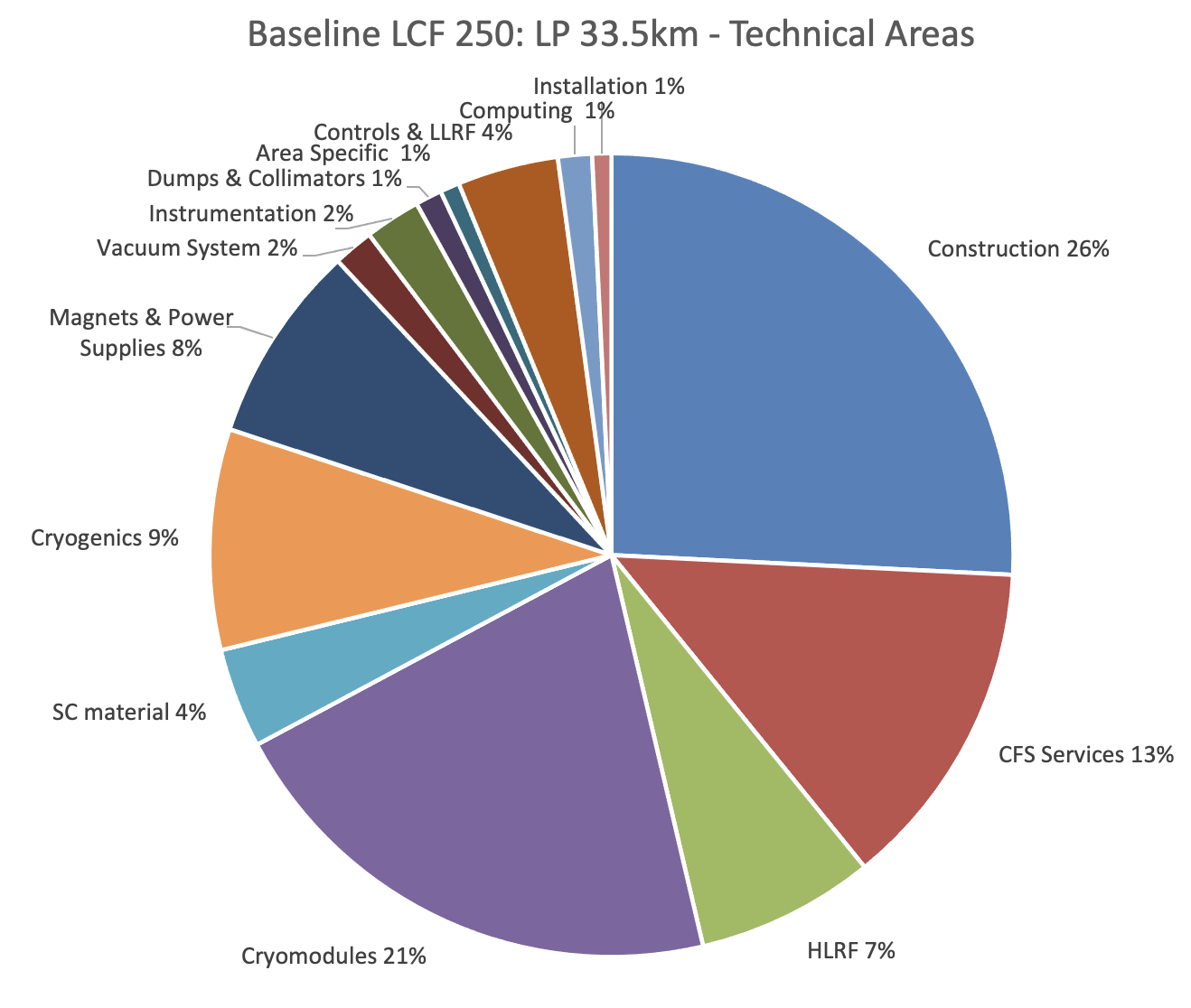}
    \includegraphics[width=0.48\hsize]{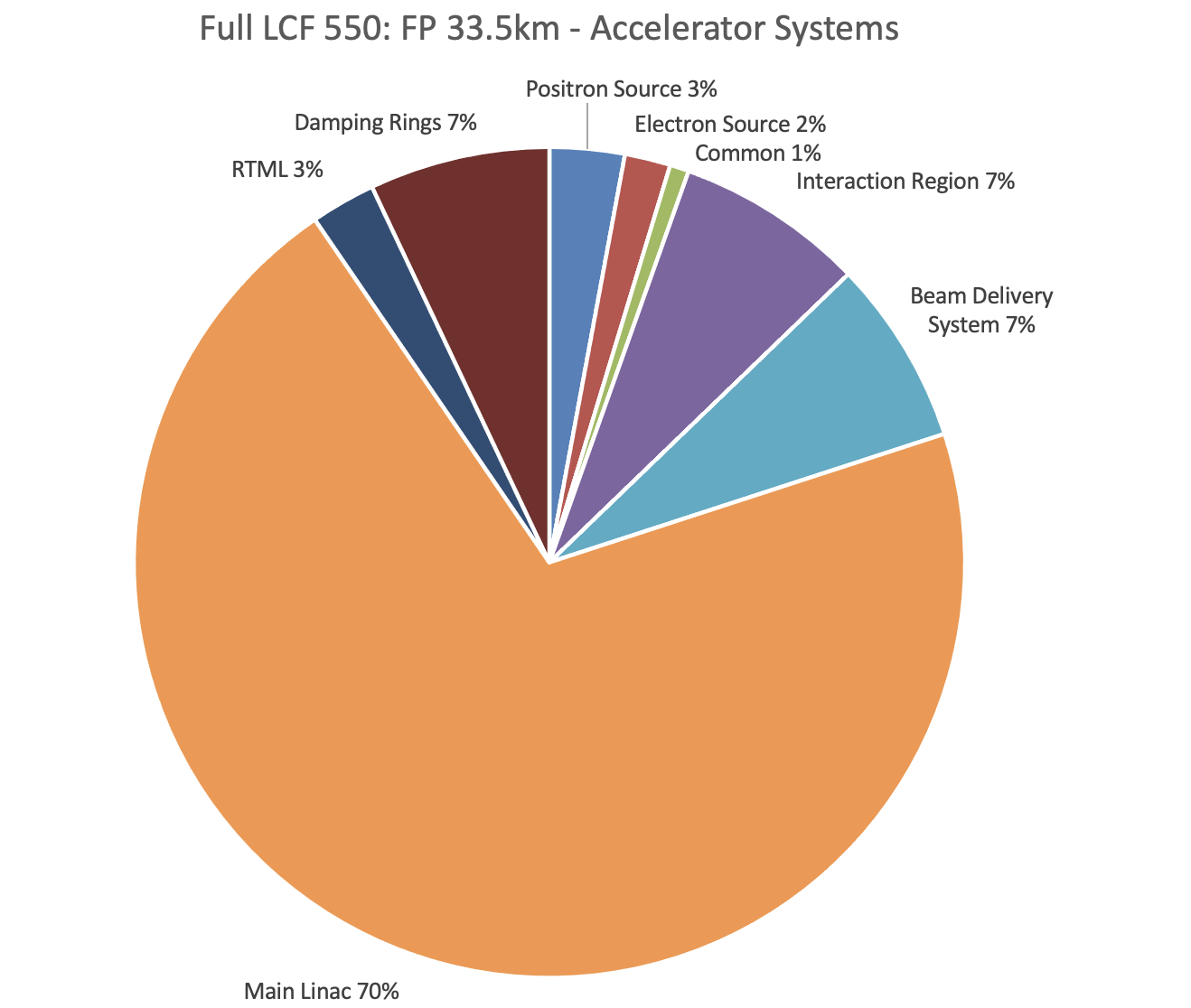}
    \includegraphics[width=0.48\hsize]{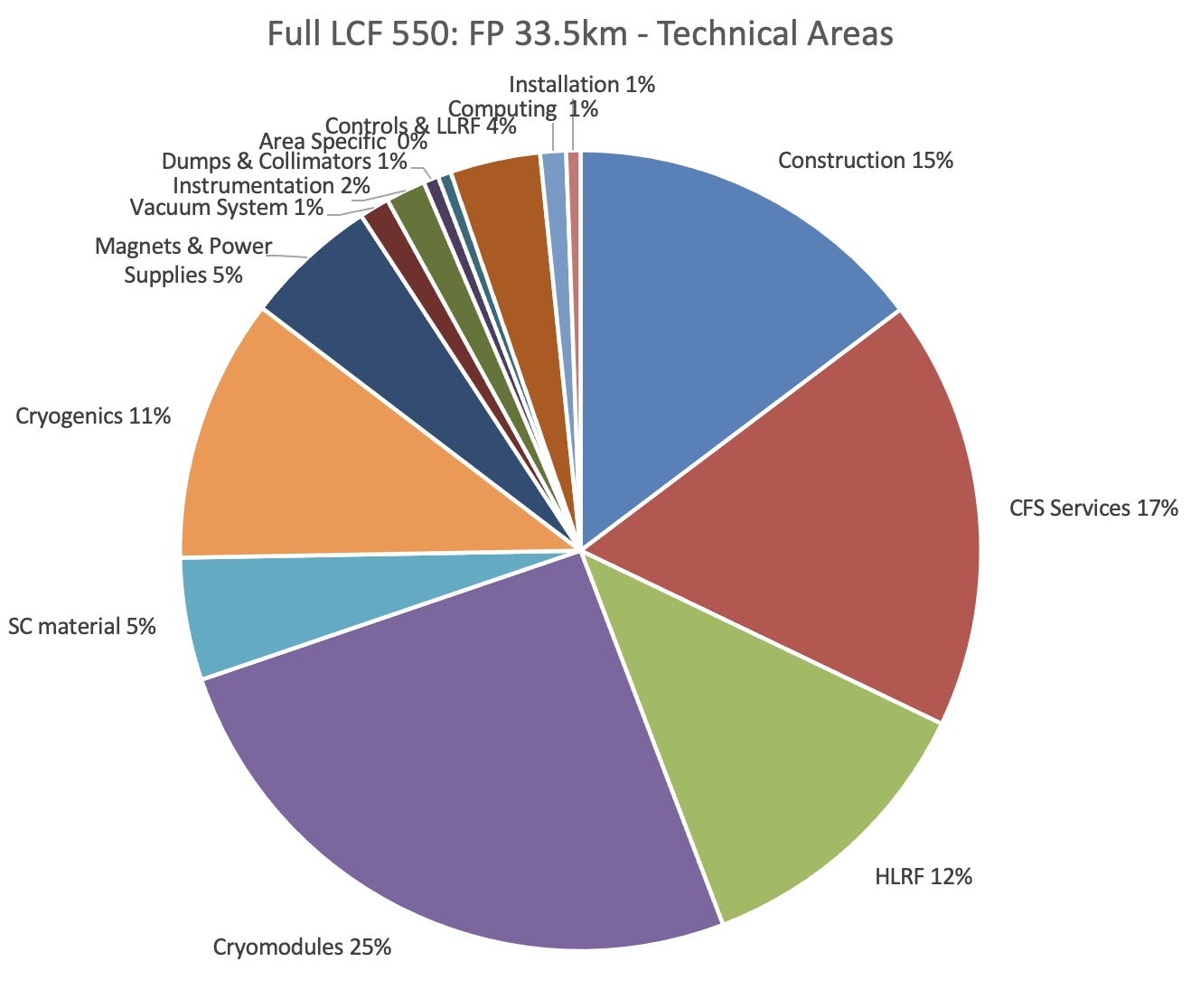}
\caption{\label{fig:costs}Costs for a Linear Collider Facility at CERN in the baseline \SI{250}{GeV} low-power (1312 bunches) configuration (top) and the final \SI{550}{GeV} full-power (2625 bunches) configuration (bottom), broken down according to accelerator areas (left) and technical systems (right).
The total cost for the baseline configuration LCF 250 LP amounts to \SI{8.29}{BCHF}, for the full configuration LCF 550 FP to \SI{14.53}{BCHF} in 2024 prices.
}
\end{figure}

\paragraph{Design differences} that were taken into account between the ILC design for Japan and the LCF design for CERN are:
\begin{itemize}
    \item a second BDS: To account for the cost of a second BDS, all costs associated with the BDS and the interaction region were doubled, with the exception of the construction costs, because those were evaluated separately to include a design for the 2nd BDS;
    \item beam energy and main linac length;
    \item cryogenic power, arising from different repetition rate and different quality factor $Q_0$;
    \item CERN being not a greenfield site: The ILC cost estimate assumes a greenfield site and thus includes cost items that are already present at CERN, in particular investment for computing (computing centre and computing capacity for accelerator simulation and operation, without the computing capacity for  the experiments). These items were removed from the cost estimate for CERN.
\end{itemize}

\paragraph{Some performance differences} between ILC and LCF are considered to arise from successful R\&D, with the assumption that these performance improvements will not affect component costs. 
These are in particular:
\begin{itemize}
    \item an increase of the quality factor $Q_0$ of the superconducting cavities from $1\times 10^{10}$ to $2\times 10^{10}$: This increase is achieved by a modification of the heat treatment recipe, which is not expected to have a sizeable cost impact, assuming that the overall yield stays the same;
    \item better klystron efficiency, increased from \SI{65}{\%} to \SI{80}{\%}: This improvement results from a change in the electron optics of the klystron, based on a better design \cite{Syratchev:2022a}.
    While this makes the klystron larger and heavier, the cost increase is hard to quantify and it is assumed that any resulting cost increase is covered by the cost uncertainty assigned to this item.
\end{itemize}

\paragraph{The cost uncertainty,} or cost premium, of the cost estimate is defined as the difference between cost estimate at \SI{50}{\%} (median) and \SI{84}{\%} confidence level, which for normally distributed data  corresponds to a $1\,\sigma$ uncertainty. 

The uncertainty associated with each cost element depends on the nature, quality and maturity of the basis of estimate for that element. 
In the TDR estimate, uncertainties were evaluated by cost estimators for each cost element, based on defined guidelines, taking into account:
\begin{itemize}
    \item the maturity of the item's design (conceptual, preliminary, or detailed);
    \item the level of technical risk involved in the design and manufacture of the item;
    \item the impact of delays in this item on the project schedule (critical-path impact, non-critical-path impact, no schedule impact on any other item);
    \item the source of the cost information (engineering estimate based on minimal experience, engineering estimate based on extensive experience, vendor quote, industrial study, catalogue price);
    \item the extent, if any, of cost scaling to large quantities.
\end{itemize}
Uncertainties from the TDR were retained; for new estimates, uncertainties are determined according to the same guidelines.

For line items whose cost estimate has not been updated in 2024, but has been escalated from 2012 prices, the uncertainty of the regional escalation factor has been evaluated, and the resulting error has been added in quadrature to the original cost premium.

Cost premiums of line items are added linearly, which corresponds to treating all uncertainties as fully correlated and is a conservative approach\footnote{Treating them as uncorrelated would lead to an unrealistic reduction of the cost uncertainty when large cost items are split into smaller ones}.

\paragraph{The cost uncertainty reflects cost risk,} i.e.\ uncertainties or errors in the cost basis (e.g.\ procurement of a similar item, quantity discount from a single unit price, engineering estimate, etc.) on which the cost of a specific item is based.
This is to be distinguished from \textbf{technical risk} that is related to failure of a specific item to achieve the design performance, requiring a redesign which may result in schedule delays and increase the cost, and from
\textbf{schedule risk} that is related to failure to supply a specific item on schedule, requiring delays which may increase the cost (typically by introducing inefficiencies and additional manpower requirements).
\textbf{Market risk} is related to deviations in procurement costs from the estimate, due to changes in economic market conditions between when the estimate was made, and when the procurement is made.
\textbf{Contingency} is a broader term, and includes not only cost uncertainties but also, for example, allowances for missing items.
Performance risks and opportunities are discussed in detail in Sec.~\ref{sec:risk}.\\
\noindent
\textbf{The resulting cost premium for the overall cost has been determined to be \SI{29}{\%}.}


\subsection{Operating costs}

\paragraph{Operational year and energy consumption}
The assumptions on the operational year are summarised in Table~\ref{tab:op-year}.
The numbers correspond to an availability of \SI{75}{\%}, in accordance with ILC design parameters.
The approach follows the one laid out in 
\cite{Bordry:2018gri} for FCC-ee and CLIC.

\paragraph{The power consumption,} relative to the power for data taking, has been calculated under the following assumptions: 
Four operating modes are assumed -- data taking, standby, shutdown, and machine development. 
In standby mode, RF is switched off, but all other systems are assumed to be operational. This saves between \SI{46}{\%} and \SI{56}{\%} of power. Standby mode is assumed to correspond to the state during down time and technical stops.
In shutdown mode, RF is switched off permanently, so that the cryogenic load is reduced to about \SI{32}{\%} of its nominal value (calculated from the ratio of static to total cryogenic load, see \cite[Tab. 3.11]{Adolphsen:2013kya})
as well as magnets and other accelerator systems are switched off. Remaining power loads are the civil infrastructure (lighting, cooling, ventilation, etc.), and cryo plants satisfying the static load. In this mode, between \SI{25}{\%} and \SI{28}{\%} of the nominal power are estimated to be consumed.
These operating scenarios are the basis for the calculation of the total yearly energy consumption, as summarised in Table~\ref{tab:ghg-project-ILC}.

\begin{table}[htbp]
\small
    \centering
    \begin{tabular}{lccccc}
         &  & \multicolumn{4}{c}{Fraction of peak power}\\
       Operational phase   & days/year & LCF 250 LP &  250 FP &  550 LP &  550 FP \\
     \hline
       Annual shutdown      & 120 & \SI{28}{\%} & \SI{18}{\%}  & \SI{28}{\%} & \SI{25}{\%} \\
       Commissioning        & 30  & \SI{77}{\%} & \SI{72}{\%}  & \SI{75}{\%} & \SI{72}{\%} \\
       Technical stops      & 10  & \SI{54}{\%} & \SI{45}{\%}  & \SI{50}{\%} & \SI{44}{\%} \\
       Machine development  & 20  & \SI{77}{\%} & \SI{72}{\%}  & \SI{75}{\%} & \SI{72}{\%} \\
       Downtime (faults)    & 46  & \SI{54}{\%} & \SI{45}{\%}  & \SI{50}{\%} & \SI{44}{\%} \\
       Data taking          & 139 & \SI{100}{\%} & \SI{100}{\%}  & \SI{100}{\%} & \SI{100}{\%} \\
    \end{tabular}
    \caption{Operational phase definition, with relative power consumption for each operation phase in relation to the nominal operating power assumed in the electricity consumption calculation.}
    \label{tab:op-year}
\end{table}

\paragraph{Operating costs} are summarised in Table~\ref{tab:op-costs}.
They were calculated taking into account the yearly electricity consumption, at a price of \SI{80}{CHF} per MWh, and the costs for spares and maintenance.

Spare and maintenance costs were evaluated in relation to the construction costs as follows:
Construction items are categorised into
\begin{itemize}
    \item fixed accelerator installation, taken to be the total costs for cryomodules, magnets, vacuum system;
    \item consumables, taken to be the costs for all other accelerator components, i.e., HLRF equipment, magnet power supplies, dumps, and cryogenics\footnote{For cryogenics, experience from LHC indicates that the yearly operating and maintenance costs are \SI{2.8}{\%} of the original invest, which includes the costs for replenishing liquid helium losses \cite{Delikaris:pc2025}.}, area-specific equipment such as source targets and guns, controls and LLRF, instrumentation, and computing;
    \item technical infrastructure, taken to be the costs for civil infrastructure such as water cooling, electric power infrastructure, air handling, installation equipment, and \SI{10}{\%} of the civil construction cost,

\end{itemize}
and assuming a replacement and maintenance cost per year of \SI{1}{\%} / \SI{3}{\%} / \SI{5}{\%}, respectively, of the corresponding capital expense.
The operating costs for cryogenics includes the costs for liquid helium losses.

\paragraph{Human resource needs for operation} have been estimated for the ILC \cite{Dugan:2013a} to be \SI{850}{FTE} for the \SI{500}{GeV} and  \SI{638}{FTE} for the \SI{250}{GeV} machine.
These numbers are consistent with estimates for CLIC \cite{Aicheler:2018arh} and the general experience at CERN.

\begin{table}[htbp]
\small
    \centering
    \begin{tabular}{lccccc}
     Operating costs                        & LCF &  250 LP &  250 FP &  550 LP &  550 FP \\
     \hline
     Nominal operating power                    & MW   & 143 & 182 & 250 & 322 \\
     Yearly electricity consumption             & TWh  & 0.8 & 1.0 & 1.4 & 1.8 \\
     Electricity price & CHF/MWh & \multicolumn{4}{c}{80} \\
     Yearly electricity costs                   & MCHF &  66 &  77 & 115 & 142 \\
     \hline
     Fixed accelerator inst. (1\,\%)            & BCHF   & 2.8 & 2.9 & 5.2 & 5.3 \\
     Consumables (3\,\%)                        & BCHF   & 2.4 & 2.8 & 4.0 &4.8 \\
     Technical infrastructure (5\,\%)           & BCHF   & 1.1 & 1.4 & 2.0 & 2.5 \\
     Operation \& maintenance                   & MCHF/y & 156 & 182 & 273 & 322 \\
     \hline
     Total yearly operation costs               & MCHF/y & 222 & 259 & 388 & 464 \\
     \hline
     Personnel                                  & FTE  & 640 & 640 & 850 & 850 \\
    \end{tabular}
    \caption{Operating costs, for electricity and operation and maintenance (including helium, see text).
    }
    \label{tab:op-costs}
\end{table}

\section{Sustainability and environmental impact}

The construction and operation of large-scale accelerator infrastructures such as the planned LCF at CERN require considerable resources and therefore entail a significant impact on the environment throughout the lifecycle of the project.
As scientists and citizens, we acknowledge our obligation to future generations to plan our projects in the most sustainable way possible.
Optimisation requires quantitative knowledge of the burdens imposed by the construction, operation and demolition of the entire facility.
To this end, the ILC and CLIC projects have commissioned two lifecycle assessment (LCA) studies with ARUP, a consultancy company \cite{clic_ilc_lca_arup,clic_ilc_lca-accel_arup}. 
These lifecycle assessments form the basis for the greenhouse gas (GHG) emission numbers summarised in Table~\ref{tab:ghg-project-ILC}.

\paragraph{A first study \cite{clic_ilc_lca_arup} quantified the environmental impact of the civil construction} of the underground tunnels, caverns and shafts of the ILC and CLIC in several configurations, as it was assumed that this would constitute the largest contribution to the overall emissions. 
The study was based on the detailed available designs of the underground installations and considered the relevant construction methods.
In particular the tunnelling methods chosen are different: the ILC in Japan will use the New Austrian Tunnelling Method (NATM), a drill--and--blast method, in a solid granite geology, while at CERN a tunnel boring machine would be used in molasse rock. 
The resulting tunnel cross sections differ substantially in shape and in the amounts of material used for grouting, lining, rock bolts, and reinforcement steel.
This is reflected in the LCA.
The numbers quoted for the LCF at CERN are based on these analyses, which is possible as the LCF's tunnel design closely follows the CLIC design.

\begin{figure}[htb]
    \centering
    \includegraphics[width=0.50\hsize]{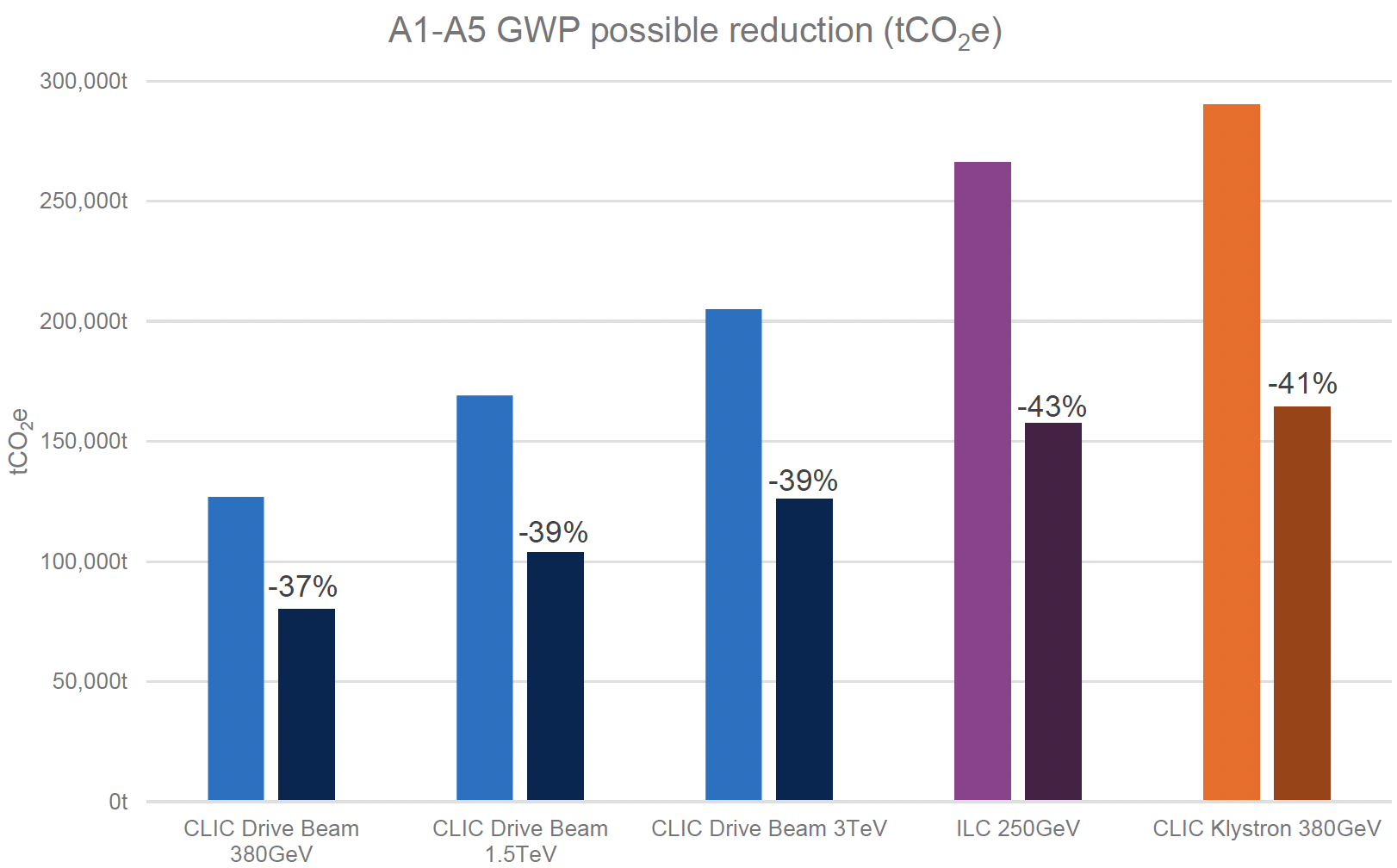}
\caption{\label{fig:co2-reduction} \ce{CO2} emission reduction opportunities for the ILC and CLIC underground facilities identified in an LCA study of underground civil construction \cite{clic_ilc_lca_arup}.  }
\end{figure}

The purpose of this and other life cycle assessment studies is not so much to produce a set of numbers that quantify the environmental impact, in particular the global warming potential (GWP) as a measure of effective greenhouse gas emissions, as to provide quantitative data early in the project to reduce the impact as far as possible. 
Important conclusions that can be drawn from the LCA study of underground facilities are:
\begin{itemize}
    \item Given the low carbon intensity of French electricity, for a project at CERN the embodied carbon in civil construction by far surpasses that arising from electricity consumption.
    \item The GHG impact of underground structures is by far dominated by the embodied carbon in the materials used for construction, in particular cement (around \SI{85}{\%}) and steel  (around \SI{15}{\%}).
    \item The GHG impact of concrete and steel can be substantially (more than \SI{40}{\%}) reduced by optimising the design (e.g.\ by a reduced lining thickness) and by usage of carbon-reduced materials, in particular as cement\footnote{For instance, usage of cement with \SI{50}{\%} addition of GGBS (ground granulated blast-furnace slag) would provide a \SI{22}{\%} overall \ce{CO2} reduction.}. The overall reduction opportunities are shown in Fig.~\ref{fig:co2-reduction}. 
\end{itemize}

\paragraph{A second, more comprehensive study \cite{clic_ilc_lca-accel_arup} also quantified the environmental impact of the accelerator and detector construction} for the CLIC and ILC designs.
Owing to the enormous complexity of an entire particle accelerator and the detectors, with a multitude of different components and materials, an LCA of a complete accelerator project is extremely challenging.
Nonetheless, an attempt was made to quantify the overall impact by focussing on the most abundant and material-intensive systems and components, in particular the magnets and accelerating RF structures.
Because some materials, in particular very pure niobium used in superconducting cavities, are specific to accelerators, so that no lifecycle data have been available in the usual databases, detailed studies of these materials were necessary as part of the overall effort. 
The same applies to specific production methods such as electropolishing used in the cavity fabrication.
This study also made an attempt, for the first time, to quantify the environmental impact of the operation phase beyond the electricity consumption (e.g.\ the impact of spare part production) and of the demolition stage  with the recycling or disposal of materials.

\begin{table}[htb]
 \footnotesize
 \centering
 \begin{tabular}{l|c|c||c|c}
   &\textbf{  ILC 250} & \textbf{ ILC 500} & \textbf{ LCF 250 LP / FP} & \textbf{ LCF 550 LP / FP} \\
 \hline
    CoM energy [GeV]                   & 250 & 500 & 250 & 550 \\ 
    Luminosity/IP [\SI{e34}{cm^{-2}s^{-1}}] & 1.35 / 2.7 & 3.6 & 2.7 / 5.4 & 3.85 / 7.7 \\ 
    Number of IPs & \multicolumn{2}{c||}{1} & \multicolumn{2}{c}{2} \\ 
    Operation time for physics/yr  $[10^7 s/ yr]$ & \multicolumn{2}{c||}{1.6}  & \multicolumn{2}{c}{1.2} \\ 
    Integrated luminosity/ yr  [1/fb/ yr] & 215 / 430 & 580 & 325 / 650 & 460 / 920 \\
    Host countries & 
           \multicolumn{2}{c||}{Japan} & 
           \multicolumn{2}{c}{France and Switzerland} \\
 \hline 
 \multicolumn{5}{l}{\textbf{GHG emissions from construction, stages A1-A5}} \\
 \hline 
    Subsurface tunnels, caverns, shafts [kt~\ce{CO2}e] & 266 & 372 & \multicolumn{2}{c}{380} \\ 
    Accelerator (coll.) [kt~\ce{CO2}e] & 150 & 270 & 165 & 310 \\
    Accelerator (inj.) [kt~\ce{CO2}e] & 59 / 82 & 83 & \multicolumn{2}{c}{60 / 84} \\
    Services [kt\ce{CO2}e] & 32 & 46 & \multicolumn{2}{c}{46} \\
    Detectors [kt~\ce{CO2}e] & \multicolumn{2}{c||}{94} & \multicolumn{2}{c}{94}\\
 \hline 
    \textbf{Total [kt~\ce{CO2}e]} & 601 / 624 & 865 & 745 / 769 & 890 / 914 \\ \hline 
    Collider tunnel length  [km] & 20.5 & 33.5 & \multicolumn{2}{c}{33.5}\\ 
    Collider tunnel diameter [m] & \multicolumn{2}{c||}{9.5}  & \multicolumn{2}{c}{5.6}\\ 
    Collider tunnel GHG / m  [t~\ce{CO2}e/ m] & \multicolumn{2}{c||}{8.6}  & \multicolumn{2}{c}{8.5}\\ 
    Concrete GHG  [kg~\ce{CO2}e/ kg] / [kg~\ce{CO2}e/ $m^{3}$] & \multicolumn{2}{c||}{0.16 / 400 (C25/30)}  & \multicolumn{2}{c}{0.16  / 400 (C25/30)} \\ 
    Main Linac accelerator  GHG / m  [t~\ce{CO2}e/ m] & \multicolumn{2}{c||}{5.3}  & \multicolumn{2}{c}{5.3}\\
 \hline 
 \multicolumn{5}{l}{\textbf{GHG emissions from operation}} \\
 \hline 
    Maximum power in operation [MW]                            & 111 / 138 & 164 & 143 / 182 & 250 / 322 \\ 
    Annual electricity consumption   [TWh/ yr]            & 0.7 / 0.9 & 1.1 & 0.8 / 1.0 & 1.4 / 1.8 \\ 
    Reference year of operation                                     & \multicolumn{2}{c||}{2040}  & \multicolumn{2}{c}{2050} \\
    Carbon intensity of electricity [g~\ce{CO2}e/ kWh]    & \multicolumn{2}{c||}{81}   & \multicolumn{2}{c}{16} \\ 
    Average Scope 2 emissions / yr [kt~\ce{CO2}e]              & 59 / 74 & 87 & 13 / 15 & 23 / 28 \\
 \end{tabular}

 \caption{Data on GHG emissions for the ILC project in Japan and the Linear Collider Facility proposal for CERN. These are baseline numbers, before application of possible \ce{CO2} reduction measures.
 The optimisation potential for tunnels, caverns and shafts is estimated to be \SI{50}{\%}, for accelerators, services and detectors it is assumed to be \SI{25}{\%}.}

 \label{tab:ghg-project-ILC}
\end{table}

Some important conclusions from this study:
\begin{itemize}
    \item The GHG impact of the accelerator and detector installations together is as big as or even bigger than the impact of civil construction.
    \item The impact is dominated by the embodied carbon in the materials used; material processing (e.g.\ refinement by electron beam melting) for specific material qualities required for accelerators (ultra pure niobium, oxygen free copper) plays an important role and cannot be neglected in the analysis.
    \item Again, significant reduction opportunities exist, but owing to the multitude of components and materials these are more diverse than in the civil construction case and require careful individual assessment.
    \item The GHG impact of electricity consumption depends on the location of the project and on the time of (first) operation owing to envisaged carbon reduction measures in the electricity sector. 
    Assumptions about these parameters have ramifications for the overall system optimisation (embodied carbon versus operation). 
    \item Detectors contribute a sizeable amount to the GWP impact, dominated by the iron required for the magnet return yokes.
\end{itemize}

The data gathered in this study were also used to estimate the \ce{CO2} impact of an LCF at CERN, based on ILC technology, as summarised in table~\ref{tab:ghg-project-ILC}.
The ILC data were adjusted, to the extent possible, for design differences, in particular a second BDS, a third damping ring for full power (2625 bunches) operation, longer tunnel lengths, and different centre-of-mass energies, to estimate the LCF numbers.

\paragraph{The peak and annually integrated energy consumption} during each stage of operation is given in Table~\ref{tab:ghg-project-ILC}, together with the integrated carbon-equivalent energy costs for the construction of the various stages. The required site AC power ranges between \SI{143}{MW} and and \SI{322}{MW} for the initial \SI{250}{GeV} and the final \SI{550}{GeV} machine. 

\paragraph{Further environmental impacts} are expected to be small in comparison. 
In particular it is assumed that direct emissions from detectors of green house gases with very high weighting factors such as chlorofluorocarbons (CFC) or \ce{SF6}, which currently dominate CERN's GHG emissions \cite{CERN:2023a}, will be fully curtailed in a future project by adequate protection measures. 
As additional information, Table~\ref{tab:ILClandoccupation} gives the estimated land use of the LCF project at CERN. It includes both CERN attributed fenced land ("fenced") and CERN
attributed unfenced land ("unfenced"). CERN buildings and infrastructure include constructions and installations,
both above and below ground. The third category is land outside the attributed areas ("outside"). For the LCF it is possible to move the IP surface infrastructure partly or fully inside the "fenced" area if this is advantageous. Re-use of
fenced land is not considered "consumed" in this summary.

\begin{table}[h!]
\footnotesize
\centering
\begin{tabular}{|l||c|c|}\hline
~ & \SI{20.5}{km} machine & \SI{33.5}{km} machine \\\hline\hline
Number of new access shafts (outside/unfenced/total) & 4 / 3 / 7 & 8 / 3 / 11 \\\hline
Number of new surface sites (outside/unfenced/total) & 4 / 1 / 5 & 8 / 1 / 9\\\hline
Area of land  permanently consumed (outside/unfenced/total)~[\SI{}{\square\km}] & 0.12 / 0.13 / 0.25 & 0.21 / 0.13 / 0.34 \\\hline
Area of surface constructions (outside/unfenced/total)~[\SI{}{\square\km}] & 0.022 / 0.033 / 0.055 & 0.040 / 0.033 / 0.073 \\ \hline
\end{tabular}
\caption{Area of land occupied by the LCF project at CERN.}
\label{tab:ILClandoccupation}
\end{table}


\section{\label{sec:techdev}Technology and delivery}





In principle, all \textbf{key technologies} for the \SI{250}{} and \SI{550}{GeV} machines \textbf{are  mature and construction-ready}. 
In particular, the superconducting radio-frequency technology is extremely well understood and at the heart of well-established and excellently performing machines like the European XFEL in Hamburg, Germany, or the LCLS~II at SLAC, USA. Still, compared to e.g.\ the ILC TDR from 2013~\cite{Adolphsen:2013kya}, significant R\&D progress has been achieved in terms of, e.g., gradients, quality factors ($Q_0$), energy efficiencies, and repetition rates (see Sec.~\ref{sec:methodology} ''Some performance differences''). A discussion of the R\&D progress and European XFEL operation experience on which the LCF capitalizes can be found in~\cite{LCVision-Generic}. 

The overall technological readiness of the project is reflected in the timeline discussed in Sec.~\ref{sec:timeline} and shown in Table~\ref{tab:timeline_ILC}. The necessary steps are given separately for the preparatory phases 1 and 2 defined in the main document of this submission.   
The following subsections~\ref{sec:rdreq} and~\ref{sec:readiness} discuss the most important remaining R\&D topics and necessary technology choices and the TRL levels of individual sub-systems of the machine, respectively.

\subsection{\label{sec:rdreq}Overview, R\&D requirements and next steps}

Despite the overall technological readiness of the project, some R\&D needs to be carried for certain aspects of an LCF machine, and important studies for siting etc.\ are required, each with significant durations. The work is split up into two preparation phases of three and five years, respectively, see Tab.~\ref{tab:timeline_ILC}. The most time-critical aspect is the exact siting -- i.e.\ a detailed placement study, which might even be technology-independent. Furthermore, also definite parameter choices (e.g.\ gradient versus number of cryomodules) need to be made. 
Both will be addressed in Phase~1,  while Phase~2 will lead up to a full CFS planning and an engineering design. 

Note that the start of the second phase -- here assumed at a time T$_0$ -- would require a formal decision to proceed with the project. Likewise, the construction start at T$_1$ requires a formal decision by CERN Council. Note further that in Tab.~\ref{tab:timeline_ILC} the construction time is increased beyond the purely technical limitation of eight years to ten years in order to accommodate the fact that the transition between phases might require some time, and also to avoid the clear conflict between HL-LHC operation and beam commissioning of a new collider. 

\paragraph{Phase~1,} ideally starting directly in 2026 after the finalisation of the EPPSU, 
builds on and integrates
the on-going work of the ILC Technology Network (ITN) on the key technologies~\cite{IDT-EB-20023-002, ILC-EPPSU:2025}, complementing it with siting and implementation studies, in parallel with design and technical studies to determine and confirm the final LCF parameters. This phase is required to prepare a project decision by CERN Council, which is not expected before 2028, and should deliver:
\begin{itemize}

\item {\bfseries Final placement of the collider complex} and its infrastructure after assessment of their territorial compatibility, both for its initial phase and potential upgrades.  Preparation and documentation for wider implementation studies with the host regions / states in Phase~2.

\item {\bfseries Optimisation of LCF design:} Review of the accelerator design along with updated cost, power and risk assessments, including in particular:
\begin{itemize}

    \item a detailed design of the Interaction Region for two experiments -- including re-optimised Beam Delivery Systems -- that is suitable for an initial SCRF-based accelerator as well as for future technology upgrades like CLIC, C$^3$, ERLs or PWA; 

    \item studies of the implications for Civil Engineering and the equipment used in the initial SCRF based facility, in order to allow collider upgrades in the future, and to accommodate the "beyond colliders" physics programme described in the main document;
    
    \item design and parameter optimisation of the entire machine including further nanobeam studies. This comprises an R\&D programme to develop prototypes of key components with regard to performance and cost risks and opportunities, or with ramifications for the CFS design, in particular cavities, cryomodules, klystrons, RF input couplers, main dumps, and the positron source target and capture device.
    
\end{itemize}

\end{itemize}

\paragraph{The estimated resources needed for Phase~1} are \SI{35}{MCHF} material and \SI{180}{FTEy} of personnel effort. These resources are backed up by the ITN efforts that, over 5 years, amount to around \SI{30}{MCHF} and about \SI{250}{FTEy}.

\paragraph{Phase~2} would require a decision to go ahead with the project preparation as needed to start construction. It targets the final engineering design, larger industrial pre-series and an extensive site preparation, in particular:

\begin{itemize}
    \item SCRF industrialisation of a design based on more ambitious goals for cavity quality factors ($Q_0$) and power (klystron) efficiency, with higher repetition rate and hence cryogenic needs than foreseen in the ILC TDR design. This task builds on the work of the ITN~\cite{IDT-EB-20023-002};
    \item detailed studies of the Cryogenics system and Infrastructure Systems (e.g.\ cooling and ventilation, electrical, access and safety systems, transport and installation) adapting them to standard solutions used by CERN and industry;
    \item final site preparation including documentation and specifications for the time-critical civil engineering contracts. 
        \item environmental studies and integration of the collider in the local area with the host states.
\end{itemize}

\paragraph{The estimated resource needs for Phase~2} are \SI{120}{MCHF} for pre-series and \SI{420}{FTEy} of personnel effort for the technical studies, pre-series, engineering design and laboratory infrastructure. 
With the widespread expertise built up worldwide for SCRF based free-electron lasers (e.g.\ the European XFEL or LCLS-II) as well as for the ILC, a significant part of this work can be done outside CERN for an LCF starting with SCRF technology. 
In parallel, civil engineering preparation including continued environmental studies will require significant resources, typically 5\% of the civil engineering budget.

Also during the preparation period, detector collaborations need to be prepared and set up, building on the experience of the numerous detector concepts for Higgs factories, but also embracing new ideas. 


\subsection{\label{sec:readiness}Assessment of technology readiness}


The technological readiness of the ILC has been scrutinised by the Implementation Task Force in the context of the Snowmass process in 2021~\cite{Roser:2022sht}, with the result that all areas considered were given the highest mark, corresponding to a Technology Readiness Level (TRL) of 7 to 8, except the positron source, which was assigned the second highest category, corresponding to TRL of 5 to 6.
Parallel to that, the IDT formulated a plan for a preparatory lab \cite{InternationalLinearColliderInternationalDevelopmentTeam:2021guz}, identifying the most important R\&D topics, based on considerations of technical risk, but also cost risks and opportunities. 
Based on a revised version of this plan \cite{IDT-EB-20023-002}, the International Technology Network (ITN) has been formed that now actively pursues these issues.



In the following, we discuss the most important R\&D topics and give our own assessment of technology readiness.
We follow the TRL definitions given in \cite[Appendix A.7]{Roser:2022sht}, which are closely aligned with the definitions in the DOE Technology Readiness Assessment Guide \cite{DOE-413.3-4A}.

\paragraph{Cryomodules and cavities} 
Cryomodules with a design that is almost identical to the design foreseen for the LCF have been successfully produced in series for the European XFEL, and very similar ones for LCLS-II, and have been in operation at the European XFEL since more than 8 years without problems.
Industrial production of cavities with more than 800 cavities produced for the European XFEL alone has been demonstrated~\cite{Reschke:2017gjp,Walker:2017a}, with many cavities surpassing ILC specifications.
For ILC specifications, the remaining risk is not any more of technological nature, but predominantly concerning the production yield.
The resulting cost risk is small, as a 90\,\% production yield is already factored into the cost calculation, and a further yield reduction of e.g.\ \SI{10}{\%} would lead to only a few percent overall project cost increase.

The operating gradient of \SI{31.5}{MV/m} required for the LCF has also been demonstrated for complete cryomodules~\cite{Broemmelsiek:2018iqr}.
All key components (couplers, cavities, tuners, cold mass and vessel) have been successfully produced in industry (\textbf{TRL 9}). The main parameter to be improved is the quality factor $Q_0$ at the operational gradient of \SI{31.5}{MV/m}, since the
LCF assumes improved performance in cavity quality factor ($Q_0=2\times 10^{10}$ instead of $1\times 10^{10}$) at an average operational gradient of \SI{31.5}{MV/m}.
Successful industrial series production of cavities with these performance parameters has to be demonstrated; with regard to the quality factor, we rate the cavities 
 \textbf{TRL 7} (prototype demonstration in operational environment). 

Overall, superconducting TESLA-type cavities and cryomodules for the main linac have been fully industrialised and thus do not constitute a technological risk for the construction of a superconducting linear collider facility as proposed, which is a main strength of the proposal.
As cavity and cryomodule production (including costs for superconducting material) constitutes 25\,\% of the overall cost of the proposed LCF, successful R\&D offers large benefits in terms of performance and construction cost, which is why this is considered a high priority item. For a more in-depth discussion of the risks and opportunities related to cryomodule performance see Sec.~\ref{sec:risk}.

The time-critical work packages WP'1 (cavity production) and WP'2 (cryomodule assembly) of the ITN \cite{IDT-EB-20023-002} have been created to conduct this R\&D.
In the context of a preparatory study for an LCF at CERN, these activities should be further intensified.

\paragraph{Positron source}
The baseline source design for the ILC as well as the LCF is the helical undulator positron source, which provides polarised positrons at a rotating target from polarised photons created in a helical undulator operated with the accelerated electron beam.
The basic principle (production of polarised positrons from photons generated in a helical undulator) has been demonstrated in a successful experiment (E166) at SLAC \cite{Alexander:2008zza,Alexander:2009nb}.

The status of the ILC positron source has been summarised by the positron working group~\cite{Yokoya:2018a}.
Since then, significant progress has been made on several fronts \cite{Moortgat-Pick:2024gbw, Moortgat-Pick:2025xve}.
This system has a number of critical key components:

The superconducting helical undulator comprises a number of modules that form a \SI{230}{m} long undulator section, creating a highly collimated, narrow beam of photons.
Helical undulators have been successfully produced, a prototype design for a cryostat as required for the ILC has been successfully built and tested~\cite{Scott:2011a}. 
The operation of long undulator sections in high-energy beams is state of the art, e.g.\ at LCLS, PAL-XFEL and European XFEL~\cite{Nuhn:2009zz,Kang:xa5013,Abeghyan:2019aa}.
A fully operational undulator module with the required parameters (k-value) would be the next step.

The design of the positron production target foresees a rotating wheel made of titanium alloy, cooled either by radiation or heat conduction via friction~\cite{Dietrich:2019rts}. 
The engineering challenges are vacuum tightness of the rotating assembly, sufficient cooling and damage from shock waves created by the pulsating photon beam.
Due to the large activation of the materials, a remotely operated target exchange mechanism is required.
First engineering prototypes of a rotating target have been built.
Target materials have been characterised in test beams, demonstrating the required radiation hardness~\cite{Lengler:2024znl}.
Overall, we consider the target design \textbf{TRL 5} (component validation), consistent with the evaluation of the ITF.
A next step will entail a fully operational target, operating at the required parameters (vacuum, rotational speed, cooling, operation in presence of pulsating magnetic field of capture device), which would raise the TRL to 6.

For the positron capturing device, several design ideas exist. 
The most promising design is based on a pulsed solenoid design~\cite{Moortgat-Pick:2024gbw}.
Pulsed solenoids have been successfully operated in other accelerator applications.
An engineering study of a pulsed solenoid is under way.
We consider the pulsed solenoid \textbf{TRL 5} (component validation). 
A next step entailing a fully operational pulsed solenoid -- operating at the required parameters (field strength, pulse shape, flat top length, aperture) -- is underway in the context of the ITN. This would raise the TRL to 6.

The photon dump, in particular its entrance window, is also challenging owing to the fact that changing the entrance point by moving the beam spot (``painting'') is not possible.  

The work packages WP'6 (rotating target), WP'7 (magnetic focusing system) and WP'11 (target maintenance) of the ITN~\cite{IDT-EB-20023-002} have been created to conduct the time-critical R\&D aspects.

In the context of a preparatory study for an LCF at CERN, these activities should be further intensified.

As a risk mitigation strategy, an alternative positron source concept based on a separate electron beam impinging on a slowly rotating target (electron-driven source) has been developed over the last decade~\cite{Nagoshi:2020blm,Kuriki:2022zjz}.
Such a source concept has important advantages in terms of ease of operation, as it decouples positron production from the electron main linac; however, it does not provide positron polarisation.
Should an undulator-driven, polarised positron source turn out to be too risky, a decision to build an electron-driven source would ensure that the project can proceed, albeit without positron polarisation.
The time-critical work packages WP'8 (rotating target), WP'9 (magnetic focusing system) and WP'10 (capture cavity and linac) of the ITN~\cite{IDT-EB-20023-002} have been created to conduct further R\&D on this concept.

\paragraph{Electron source photo injector gun}
High voltage photo guns with the characteristics required for the LCF have been built and successfully operated in accelerators, thus this system does not pose a technological risk.
In order to take advantage of experience accumulated over the last decade, and to preserve the existing know-how, the time-critical work packages WP'4 (higher voltage ILC photo-gun R\&D) of the ITN~\cite{IDT-EB-20023-002} has been created to continue these activities.

\paragraph{IP spot size / stability} 
Focussing the beams to the required nanometre size and bringing them to stable collision is the core challenge of the nanobeam technology that makes any linear collider possible.
This has been the topic of extensive R\&D over the last decades, with dedicated test facilities where the necessary technologies can be tested and verified.
The Accelerator Test Facility ATF2~\cite{Aryshev:2020a,faus-golfe:ipac2024-tupc04} is the main workhorse for this R\&D.
At ATF2, the final focus concept of the ILC, based on local chromatic correction, has been verified~\cite{Patecki:2016rgi}, together with the necessary beam stabilisation with the FONT feedback system~\cite{Apsimon:2018bpq,Bett:2022yyg}. Based on the extensive testing at the ATF2, we rate the issue of IP spot size and stability to be \textbf{TRL 6} (technology demonstration in a relevant environment).

The time-critical work packages WP'15 (final focus) and WP'16 (final doublet) of the ITN~\cite{IDT-EB-20023-002} have been created to conduct further R\&D.

\paragraph{Crab cavity} 
A number of competing designs for the crab cavities around the interaction point have been put forward and assessed in a dedicated workshop~\cite{Crab-cavity-design-review}, with a down-select of the two most promising designs for further R\&D.

The time-critical work packages WP'3 (crab cavity) has been created to conduct the necessary R\&D.

\paragraph{Damping rings} 
The ILC damping ring design is similar to planned and existing storage rings for synchrotron radiation production.
Key parameters are the same as those of operating facilities: 
with a \SI{3.2}{km} circumference at \SI{5}{GeV} beam energy, it is similar to PETRA~III, and the  normalised emittance $\gamma \varepsilon_{x/y} =\SI{4}{\mu m} / \SI{20}{nm}$ has been achieved in operating light sources.
Important challenges are the 400/800\,mA beam current for electrons/positrons, which were tested at the CESR-TA test facility~\cite{Adolphsen:2013jya}, the large dynamical aperture required for positron injection, and the small transverse (longitudinal) damping time of 13\,(7)\,ms, to be provided by radiation damping from superconducting undulators.
Lattices that fulfil the design requirements have been developed.
Overall, we consider the damping ring system design to be \textbf{TRL 7} (prototype demonstration in operational environment), in agreement with the assessment by the ITF.

Further R\&D concerning the overall damping ring system design is being conducted as ITN time-critical work package WP'12~\cite{IDT-EB-20023-002}. 

\paragraph{Injection and extraction kickers} Injection into and extraction from the damping rings requires kickers that are capable of kicking individual bunches at a minimal bunch separation of \SI{3.2}{ns}. 
The damping ring bunch pattern consists of mini trains of about 25 consecutive bunches, followed by a gap of about 10 empty bunches. 
Bunches are extracted from the tail of these mini trains, therefore the rise time is the bigger concern.
A prototype kicker has been tested successfully at ATF2~\cite{Naito:2011zz}.
Therefore, we rate the kicker system to be  \textbf{TRL 7} (prototype demonstration in operational environment), in agreement with the assessment by the ITF.

Further R\&D into the design of the kickers is being conducted as ITN time-critical work package WP'14~\cite{IDT-EB-20023-002}. 

\paragraph{Emittance preservation} 
Preserving the very small emittance of the beam exiting the damping rings along the low energy (\SI{5}{GeV}) transport line~\cite{Tenenbaum:2007zz,Kubo:2007b}, the bunch compressors~\cite{Kim:2009zzc} and the main linac~\cite{Eliasson:2006ak,Ranjan:2006a} is a central element to achieving the required small beam spot sizes and thus the target luminosity, and has been extensively studied.
An intrinsic advantage for emittance preservation are the large apertures of the accelerating L-band cavities, which lead to modest wakefield effects compared to accelerators operating at higher frequencies.

Therefore, we rate the emittance preservation to be \textbf{TRL 8} (system completed and qualified through test and
demonstration), in agreement with the assessment by the ITF.

\paragraph{Main dump}
The ILC main dump is designed for up to \SI{17}{MW} of beam power (for a single beam, sufficient for the \SI{1}{TeV} upgrade of ILC), which is more than sufficient for the full LCF 550 configuration, where the maximum \SI{23}{MW} beam power would be shared by two beam delivery systems and main dumps.
The main dump design is based on a water dump, which follows the design of the SLAC \SI{2}{MW} water beam dump that has been successfully operated in the past.
An engineering design is required for a detailed planning of the enclosing cavern, including cooling, ventilation and treatment of generated isotopes, in particular tritium. 
The entry window of the dump is another important challenge.

The design of main dump, including prototyping of the entry window, is the topic of the ITN time-critical work package WP'17~\cite{IDT-EB-20023-002}.

Based on the fact that at SLAC a water dump was build and operated successfully, albeit at a lower power rating, we consider the main dump technological readiness to be \textbf{TRL 5} (component validation).

\paragraph{RF systems} The RF systems (modulators, klystrons, power distribution system) required to operate the accelerator have parameters that are mostly identical to those of the corresponding systems of operational accelerators, in particular the European XFEL.

Klystrons with the required parameters (pulse length and output power) are commercially available and are in use in comparable facilities (TRL 9). 
The LCF proposal assumes that the efficiency can be raised from 65\,\% (commercially available) to 80\,\%, based on a design study for \SI{1.3}{GHz} and experience from similar developments~\cite{Syratchev:2022a}. 
Because no prototypes for the specific operation parameters (\SI{1.3}{GHz}, \SI{10}{MW}, \SI{1.65}{ms}) have been built, we rate this as \textbf{TRL 4} (validated in laboratory environment).
It is worthwhile to note that the klystron efficiency does not pose a technical risk to the LCF. 
If the expected efficiency improvement would fail to fully materialise, this would be know early enough to install more powerful modulators\footnote{Cost savings from reduced modulator capacity are not taken into account in the LCF cost estimate, which assumes \SI{65}{\%} in the costing, thus an \SI{80}{\%} is a cost reduction opportunity.} and cooling systems. 
The resulting impact on construction and operation costs is in the few percent range.

A prototype of a design suitable for installation and operation in the accelerator tunnel with the required availability parameters would be beneficial. 
The modulator foreseen for the ILC (semi\-conduc\-tor-based Marx modulator design) is different from the modulator employed in the European XFEL (bouncer modulator and pulse transformer), operating at the full high voltage of the klystron without the need for a transformer.
Prototype Marx modulators exist and have been tested, and we consider this \textbf{TRL 7} (prototype demonstration in operational environment), in agreement with the assessment by the ITF.

The waveguide power distribution system, although different in design from the European XFEL one, is comprised of commercially available components, i.e.\ \textbf{TRL 9} (proven through successful operations).

The main parameter to be improved is the klystron efficiency (\SI{65}{\%} to \SI{80}{\%}).

\paragraph{HOM detuning / damping}
Absorption of higher order modes in SRF cavities of TESLA design uses the same design employed at the European XFEL.
The components are commercially available from industry, the design has been proven at the European XFEL.
We rate this topic to be \textbf{TRL 9} (proven through successful operations).


\begin{table}[h!]
\centering
\scriptsize
\begin{threeparttable}
\begin{tabular}{|l|c|c|c|c|c|c|}
 \hline
 \multirow{3}{*}{Component/Sub-system} & \multirow{3}{*}{TRL} & \multirow{2}{*}{Main parameter}   & \multirow{2}{*}{Improvement} & \multicolumn{3}{c|}{~RF effort}  \\ 
 ~&  & \multirow{2}{*}{to be improved} & \multirow{2}{*}{factor} & Personnel  & Material  & Timescale \\
~ & ~ &  & ~ & [FTEy] & [MCHF] &  [years] \\ \hline\hline

\multirow{3}{*}{Positron source}  &  & \multirow{1}{*}{Target cooling} & \multirow{3}{*}{N/A} & \multirow{3}{*}{15} &\multirow{3}{*}{5} & \multirow{3}{*}{3} \\ 
& 5 & Pulsed solenoid & & & & \\
 &  & peak field & & & & \\ \hline
Main dump & 5 & Maximum power & 8 & N/A & N/A & N/A \\ \hline
IP spot size/stability  & 6 & Vert. beam size \@at nom. bunch population & 10~\% & 18 & 4 & 6\\ \hline
RF power sources & 6 & efficiency  & 1.2 & $ $10 & 3  & 3 \\ \hline

\SI{1.3}{\GHz} RF cavities/cryomodules  & 7 & $\mathrm{Q_0}$ & 2 & 30 & 10 & 3 \\ \hline

Damping rings  & 7 & Beam dynamics & N/A & 10 & 3 & 3 \\ \hline

 \end{tabular}
\end{threeparttable}
  \caption{Technical readiness and R\&D requirements for LCF@CERN. 
}
 \label{tab:LCFTRL}
 \end{table}


\paragraph{All further stages,} i.e.\ going to \SI{1}{TeV} or beyond, require significant R\&D. However, since these stages will enter any construction phase at the earliest in the middle of the 2050s, no critical paths or R\&D needs can be discussed here in a meaningful way. 


\section{Dependencies}

The described scenarios assume \textbf{CERN as a host site} for the Linear Collider Facility. However, the construction of an ILC-like machine has also been studied, naturally, for Japan, and other locations. 

With its modest power needs and operation costs compared to other projects, the Linear Collider Facility in the stages up to \SI{550}{GeV} centre-of-mass energy would not interfere negatively with other activities on the CERN site. In particular, a \textbf{rich scientific diversity programme} and a broad portfolio of advanced accelerator R\&D activities (e.g.\ magnet developments, plasma, ...) could be maintained as an important complement to CERN's portfolio, as postulated also by the 2020 European strategy update.

The LCF realisation and operation at CERN does, in particular, not depend on the LHC injector chain and could thus be developed independently of the ongoing (HL-)LHC programme. 

Furthermore, the realisation of the LCF programme at CERN does not entail the premature definition of the precise technology for a future \SI{10}{TeV} parton-parton centre-of-mass energy collider.

\section{Commentary on current project status}

The ILC project has, since decades, assembled a \textbf{large community} supporting it and bringing it to the level of maturity that is observed today. This is also documented in the project TDR~\cite{Behnke:2013xla,Behnke:2013lya}, in the ILC staging report from 2017~\cite{Evans:2017rvt}, and e.g.\ in the report to the Snowmass process 2021~\cite{ILCInternationalDevelopmentTeam:2022izu}. The ILC has been, in fact a \textbf{global project} for 20 years, with accompanying technology collaborations like the TESLA Technology Collaboration, TTC, for the advancement of SRF technology. 

The \textbf{size of the community} of scientists who directly work on the project can be estimated from the number of institutions supporting the TDR (rougly 400) or the number of individuals that signed the Snowmass report (rougly 500). 
The relevant community event -- the International Linear Collider Workshops (LCWS) -- typically attract between 300 and 800 participants. 

Another measure for the potential size of potential community is the expected \textbf{number of collaborators} for a facility with two large collider experiments and an ample and diverse "beyond colliders" programme attached. Assuming the typical size of an experimental collaboration of 3,000 members, and 1,000 members for the diversity programme, the Linear Collider Facility might be foreseen to hold a community of around 7,000 members. This leaves room for significant activities on continued HL-LHC data analysis, small and medium size experiments at CERN and national labs as well as intensified R\&D towards the exploration of the 10-TeV scale and beyond.

The project in both scenarios A and B lends itself easily to the realisation of significant \textbf{in-kind contributions}. 
In particular the mass production of cryomodules -- as already exercised in an industrialised fashion e.g.\ for the European XFEL -- allows for distributed the production of attractive technology in different world regions. 
Several cavity vendors take a keen interest in the project, and the highly modularised production possibilities allow for a high added economic value.   

\section{Performance Risks and Opportunities}
\label{sec:risk}
In this section, we discuss the most prominent risks and opportunities, mitigation strategies and possible impacts on cost, schedule and performance.
We highlight risks and opportunities associated with superconducting RF technology, which have the potential to have the largest cost impact owing to the large cost fraction of the main linac SCRF components.

\subsection{Accelerating gradient and collision energy}

The operating gradient of the main linac cryomodules is directly linked to the achievable beam energy and thus the physics performance.
For a Higgs factory with \SI{250}{GeV} design energy, operation at an energy reduced by 8\,\% to \SI{230}{GeV} would not compromise physics performance too much (e.g.\ $\PZ\PH$ cross-section reduced by about \SI{15}{\%}) and thus would be acceptable initially.
An undulator-based positron source, however, the yield falls off steeply at decreased electron beam energy, which raises the importance of achieving 95\,\% or more of the design energy at least for the electron beam from the start of operation.
Thus in this section, we discuss risks and opportunities associated with the operating cavity gradient.

The baseline design of the ILC and the LCF assumes an average cavity gradient of \SI{35}{MV/m} with a \SI{20}{\%} spread for cavities measured individually in a vertical test stand prior to integration in the cyro module, and a \SI{31.5}{MV/m} average gradient for operation with beam.
These are the same values as assumed in the ILC TDR.
Experience from series production of cavities and cryomodules, in particular for the European XFEL (Eu.XFEL), and the ensuing operating experience, corroborate these values, and point to remaining challenges and opportunities.

\paragraph{Single cavity performance in vertical tests:} 
The gradient performance of single cavities as received from vendors depends critically on the production and surface treatment process, and a rigorous quality control.
The Eu.XFEL production model was build-to-print, where the vendor guarantees to follow the agreed production process rigorously, but is not required to assert a specific gradient. 
This model requires a close quality monitoring between vendor and customer during production, and permits a substantial cost benefit due to the reduced risk the vendor has to carry.
Over all, 800 cavities were procured for the Eu.XFEL from two separate vendors, one of those (RI) applied the same surface treatment protocol as foreseen for the ILC, so that the performance of the cavities from this vendor provides an excellent sample for the expected performance of industrial cavity production, despite the fact that Eu.XFEL design gradient at \SI{23.6}{MV/m} was substantially lower than the ILC design gradient.
The performance of a cavity depends critically on surface quality and can be severely deteriorated by small defects, which can often be removed by a second or possibly third surface treatment, which can be a full chemical treatment (electropolishing) at the vendor, or a much simpler and cheaper high-pressure water rinse in the testing institution.
Therefore, the protocol for acceptance or re-treatment of cavities has a decisive impact on the performance of the finally accepted sample, which needs to be taken into account when interpreting the Eu.XFEL results and calculating the expected performance resulting from the acceptance criteria for ILC, which ask for at least \SI{28}{MV/m} in vertical tests.
The results for the cavity production have been published \cite{Singer:2016fbf,Reschke:2017gjp}: 
for the company (RI) that followed the ILC surface treatment recipe, the maximum gradient (gradient limited by quench or excessive field emission indicated by X-ray emissions) for cavities as received from the vendor was \SI{33.0}{MV/m}, with a RMS spread of \SI{6.5}{MV/m} (\SI{20}{\%}), already very close to the ILC specification. 
The ``usable'' gradient, which requires in addition that $Q_0$ is better than the design value of \num{1e10}, was \SI{29.0}{MV/m}, with a RMS spread of \SI{7.9}{MV/m} (\SI{27}{\%}).
Cavities were typically re-treated up to two times (mostly by high pressure water rinse) if the gradient was below \SI{20.0}{MV/m}; 
for these cavities, high pressure rinsing typically improved the gradient by \SI{8}{MV/m}.

After following the Eu.XFEL retreatment protocols, the average maximum / usable gradient of cavities from RI was \num{34.6} / \SI{31.2}{MV/m}, averaged over both vendors the average usable gradient was \SI{29.9}{MV/m} with a \SI{5.2}{MV/m} (\SI{17}{\%}) spread.

Cavities with gradients above the XFEL acceptance limit of \SI{20}{MV/m}, but below the ILC limit of \SI{28}{MV/m} were not retreated. 
An simulation of the effect of such a re-treatment regime \cite{Walker:2017a,LCVision-Generic} concluded that with up to two re-treatments, the ILC gradient goal (\SI{35}{MV/m} on average, at least \SI{28}{MV/m}) would be achieved for the usable gradient with a yield of \SI{91}{\%}, fulfilling the \SI{90}{\%} yield assumption of the TDR that underlies the cost estimate.
Thus, the Eu.XFEL production data fully validates the performance requirements for cavities after the vertical tests assumed in the ILC design, for an ILC specificatiom of $Q_0>\num{1e10}$.
The implications of the larger $Q_0$ requirement for the LCF will be discussed in the next section.

\paragraph{Performance in cryomodule at the test stand:} 
During the Eu.XFEL production, cavity performance was measured again after installation in cryomodules for \SI{100}{\%} of the modules \cite{Kasprzak:2018kkr}.
The results differ from the vertical test stand results in several ways: 
Contamination of the cavity during assembly, or relocation of dust particles when the cavity is measure in a horizontal rather than vertical position, may result in a deterioration of the achievable gradient.
Differences in the detection of X-rays signalling field emission may result in changed limits from FE.
$Q_0$ is not measured within the cryomodule, therefore the module tests results correspond more to the maximum than the ``usable'' gradient definition.
Also, gradient measurements within Eu.XFEL modules were limited to \SI{31.0}{MV/m} due to RF power limitations and to avoid any damage to cavities or input couplers, as no operation of cavities at Eu.XFEL at such high gradients is foreseen.

Comparing the measured average gradient in the cryomodule, which is cut off at \SI{31.0}{MV/m}, to the vertical test results clipped to the same gradient, shows a \SI{5.3}{\%} gradient reduction to \num{28.5} from \SI{30.1}{MV/m} for the maximum gradient and a \SI{3.8}{\%} gradient reduction to \num{27.5} from \SI{28.6}{MV/m} comparing the operational to the usable gradient \cite{Kasprzak:2018kkr}.

Cryomodule production data show that the performance difference between vertical test and module tests was larger for the first half of modules produced than for the second half.
This correlates with a significantly improved production rate, where towards the end of production 1.25 modules were produced every week, i.e., one module every four days, more than the original target rate of 1 module per week. 
Numerous improvements were made during the course of cryomodule series production, resulting in faster turn around and better quality, as expected from learning curves.

\paragraph{Consequences of gradient spread:} 
In order to increase the yield of usable cavities, a spread in usable gradient of $\pm \SI{20}{\%}$ around the mean value is accepted, for Eu.XFEL as well as ILC and LCF.
This has important ramifications for operation: one klystron supplies \num{32} cavities at Eu.XFEL and \num{26} (full power) or \num{39} (low power) at ILC/LCF, and the RF power has to be distributed such that during operation cavities with reduced gradient do not surpass their gradient limit, while better performing cavities receive enough power to reach their operating gradient.
At Eu.XFEL, the waveguide distribution system contains asymmetric shunt tees that are tailored prior to installation based on the cryomodule test results to achieve the desired RF distribution \cite{Choroba:2017ohi}.
In contrast, the ILC design foresees remote controlled variable power splitters \cite[Sect. 3.6.4]{Adolphsen:2013kya}, which are more expensive but add operational flexibility, making it possible to adjust the RF distribution based on the observed cavity performance in the beam.
Thus, at ILC one can react to any changes in gradient limitations of individual cavities. 
This can be gradient reductions that would, for a fixed distribution, force a gradient reduction for a full klystron or detuning of the cavity, or improvements, e.g. from conditioning of the cavities.
This design makes it possible to reduce the fraction of cryomodules tested from  \SI{100}{\%} to \SI{33}{\%} once stable production performance has been established, and it will reduce the difference between gradients measured in the cryomodule tests and those achieved in actual beam operation.

The use of variable splitters is a mitigating measure to reduce performance risks arising from cavity performance degradation between cryomodule production and testing and actual operation with beam.

\paragraph{Performance in beam operation:}
As the largest operating accelerator facility based on TESLA technology, the European XFEL went into operation in 2017.
It reached design energy on year later  \cite{Kostin:2019uib}, despite the fact that only 97 cryomodules out of the planned 101 modules had been installed and 21 cavities (\SI{2.7}{\%}) of cavities were detuned, i.e., did not contribute to acceleration.
12 of the detuned cavities were functional, but at a reduced level, so that the performance penalty of detuning them was lower than operating all 32 cavities powered by the same klystron at reduced gradient.
In the ILC design with variable splitters, these cavities could likely have been operated.
Overall, the cavities reached \SI{93.6}{\%} of the performance measured in the  module tests \cite{Kostin:2019uib}.
Comparisons between module test results and operation performance in beam show that on a cavity-by-cavity basis there is a sizeable (several \SI{}{MV/m}) random difference between both results \cite{Kostin:2019uib}. 
In the fixed, tailored power distribution system of Eu.XFEL the weakest cavity limits the operation of 32 cavities, costing several \SI{}{MV/m}, an effect avoided by the ILC design.

The Eu.XFEL has remained in steady and successful operation, with excellent operating experience~\cite{Schmidt:2023cqp}, see also Sec.~4.1.2 of~\cite{LCVision-Generic}.
It is due for its first warm up in the second half of 2025.

Operational experience from Eu.XFEL indicates that achieving an operational gradient in beam of \SI{31.5}{MV/m}, i.e.\ \SI{90}{\%} of the gradient measured in vertical tests (\SI{5}{MV/m}) is challenging.
Approximately \SI{5}{\%} gradient reduction are necessary as operation margin \cite{Walker:pc2025}, thus only \SI{5}{\%} additional gradient reduction are acceptable.

\paragraph{Risk mitigation at LCF:}
The mitigation strategy of the LCF to cope with a situation where less than the nominal gradient is provided by the cavities has several components, and takes into account the differences in severity, depending on the actual amount of missing gradient.

For a Higgs factory with a nominal \SI{250}{GeV} centre-of-mass energy, operation at \SI{230}{GeV} (at a \SI{8}{\%} reduction of the gradient) would still yield valuable physics results.
The yield characteristics of the undulator-based positron source, however, make it necessary to achieve at least \SI{120}{GeV} beam energy for the electron beam, to avoid a severe impact on the physics programme.
Therefore, at least \SI{96}{\%} of the nominal beam energy need to be available after commissioning to start physics operation, while it would be acceptable if it takes a couple of years (including the opportunity for technical interventions during shut downs)  to reach \SI{100}{\%}.

The ILC and LCF baseline designs for \SI{250}{GeV} foresee the installation of 837 cryomodules in the main linac, 423 (414) for the electron (positron) linac.
Starting with an energy of \SI{15}{GeV} after the two stage bunchg compressor, at the nominal gradient of \SI{31.5}{MV/m} this would provide an end energy of \num{134.4} / \SI{131.9}{GeV}; the larger energy in the electron linac accounts for the \SI{3}{GeV} energy loss in the undulator for the positron source.
Thus, the main linacs are designed with a \SI{7}{GeV} or \SI{6}{\%} reserve.
Operation of the main linac modules at \SI{28.5}{MV/m} would thus result in beam energies of \num{123.0} / \SI{120.8}{GeV}, enough to provide \SI{240}{GeV} centre-of-mass energy and operate the positron source.

The LCF design for \SI{550}{GeV} foresees 1881 (954 + 927) cryomodules, providing \num{284.3} / \SI{276.7}{GeV} of beam energy. 
There is space for the installation of at least 9 + 45 more cryomodules, enough for \SI{570}{GeV} centre-of-mass energy at nominal gradient.

The nominal operating gradient of \SI{31.5}{MV/m} is the central estimate for the realistically achievable gradient in operation at LCF. 
A reserve of \SI{6}{\%} additional cryomodules is included in the design, and initial operation at \SI{4}{\%} lower gradient would be acceptable for physics operation, providing close to \SI{10}{\%} gradient reserve.
The initial \SI{250}{GeV} stage of LCF has ample tunnel space to raise the number of installed cryomodules, which makes it possible to revise the design even during the cryomodule production stage, after the performance of modules from series production has been measured.

This the risk mitigation strategy concerning the operational gradient consist of a \SI{6}{\%} reserve of installed cryomodules, and the option to re-assess the most cost-effective combination of operating gradient and number of installed cryomodules until late in the production and installation stage of the project, with small cost and schedule risk.   

\paragraph{Cost impact of risk, cost opportunities:}
The assumption of a higher gradient for single cavities in vertical tests, compared to the Eu.XFEL production, is justified by a cavity overproduction of \SI{11}{\%}, requiring only \SI{90}{\%} yield, in contrast to a \SI{0}{\%} overproduction at Eu.XFEL, in conjunction with harder criteria for the re-treatment of cavities. 
This overproduction and the number of additional re-treatments and tests is included in the cost estimate.
A future re-assessment of the expected gradient performance and yield might result in a revision of these parameters. 
Changing the yield from \SI{90}{\%} to \SI{80}{\%}, for instance, would increase the cavity production cost by \SI{13}{\%}. 
As cavities (including material) account for approximately half of the cryomodule costs, and cryomodules for a quarter of the project cost of LCF250, this would result in an overall project cost increase of less than \SI{2}{\%}.
This cost risk is balanced by cost opportunities of comparable size: Eu.XFEL production data shows that a sizeable fraction of cavities shows gradient of \SI{40}{MV/m} and more, thus the production yield could well be better than \SI{90}{\%}; further cost opportunities are given by the reduction of electro-polishing treatments from new surface treatments and because Eu.XFEL production experience shows that for most of re-treatments a cheap high-pressure rinse is sufficient instead of the EP step assumed in the costing.

A possible re-assessment of the realistic operating gradient, assumed to be \SI{31.5}{MV/m} in the baseline design, would entail installation of more cryomodules together with more klystrons and modulators, and require and increased cryogenics capacity. 
Still, the complete SRF costs including cryogenics and HLRF entail \SI{40}{\%} of the total project cost, and thus a \SI{10}{\%} change of the assessed operating gradient would entail only a \SI{4}{\%} cost increase for the project overall.

\subsection{Quality factor $Q_0$ and cryogenic power, new surface treatments}

The ILC design parameters of \SI{35}{MV/m} and $Q_0=\num{1E10}$ in vertical tests have been the baseline since the publication of the TDR~\cite{Adolphsen:2013kya}; the staging report of 2017~\cite{Evans:2017rvt} formulated a more optimistic R\&D goal,  raising the gradient by \SI{10}{\%} to \SI{38.5}{MV/m} (\SI{35}{MV/m} in beam) and increasing $Q_0$ at the same time to  \num{2E10} (\num{1.6E10} in beam).
Since the publication of these reports, significant progress has been made towards higher gradients and higher quality factors \cite[Sect. 4.1.1]{LCVision-Generic}.
In particular, the so--called mid--T bake surface treatment \cite{Posen_mid-T_baking} followed by a low--T bake are promising, with single cell test results that already meet these targets  \cite{Steder}, and successful industrialisation of these treatments in full cryomodules for SHINE~\cite{Pan:PRAB2024}.
In addition, mid--T / low--T baking offers further cost reduction opportunities, as it removes the need for a second electro-polishing (EP) step after the heat treatment, saving a cost intensive production step with a high environmental impact.

This progress is reflected in an LCF parameter set that assumes an quality factor of $Q_0=\num{2E10}$, twice higher than the ILC baseline, at an unchanged target gradient, as a realistic estimate of the future state of the art.
The 2017 R\&D goal of achieving in addition gradient increase constitutes a further cost reduction opportunity beyond this central value.

In contrast to the onset of field emission or breakdown, gradient limitations due to $Q_0$ as included in the ``usable'' gradient definition of Eu.XFEL \cite{Reschke:2017gjp} does not constitute a hard limit for cavity operation.
In fact, in module tests $Q_0$ is not even measured, and operating gradients are set by the hard limits \cite{Choroba:2017ohi}, where $Q_0$ only enters if the degradation is so severe that a coupler power limit is reached.

Insofar, the nominal $Q_0$ target is predominantly a parameter that affects the estimate of the necessary cryogenic cooling capacity and the costs associated with this.
The best estimate for a realistic performance can and will be reassessed after the R\&D phase, at the start of procurement, and in the production phase when actual performance results from series production are available.
Changes in the assessed average $Q_0$ value could result in changes of cryogenic plant dimensions or in the target operating gradient, or a combination of both.

The cost risk of changes to the cryogenic system is small: 
only $1/3$ of the cryogenic power scales with $1/Q_0$, namely the losses in the cavity walls, while $2/3$ are from other losses (static, losses in input couplers and HOM absorbers), and cryogenics contribute less than \SI{10}{\%} to the overall project cost for LCF250LP, ergo changing $Q_0$ back to \num{1E10} would have only a \SI{3}{\%} impact on overall project costs.

Overall, the risk associated with the $Q_0$ target is acceptable, and balanced by corresponding opportunities.
The mitigation strategy in the design and production phase is a possible revision of gradient targets and/or cryo plant dimensions, with a small associated cost risk. 
In the operation phase, any unforeseen degradation of $Q_0$ would lead to a reduction of the pulse repetition frequency, associated with a reduction in luminosity, which is an acceptable performance risk that does not endanger the physics goals of the LCF.

\subsection{Risk assessment conclusion}
In summary, we assess that the ILC target gradient target for vertical tests of  \SI{35}{MV/m} at $Q_0 = \num{1E10}$ at a \SI{90}{\%} production yield, on which the cost estimate is based, has been demonstrated by Eu.XFEL production data to be the current state of the art of commercially available cavities.
The LCF target of $Q_0 = \num{2E10}$ at the same gradient, production yield and cost is a realistic goal, based on R\&D results for novel surface treatments, with significant opportunities for further performance increase and/or cost reduction.

The nominal target gradient of \SI{31.5}{MV/m} for beam operation (\SI{90}{\%} of the gradient in vertical tests), with at least  \SI{28.5}{MV/m} of gradient available after first commissioning, while not demonstrated in operational facilities, is a realistic performance goal. 
Equipping the RF distribution with variable power splitters will reduce the performance gap between module test and beam operation.
A targeted R\&D effort, starting with a careful analysis of experience from operating facilities such as Eu.XFEL, LCLS-II and others, will be required to deliver further improvements.

The performance goals given constitute our current best assessment of the performance realistically achievable by the time the project will be implemented, balancing remaining performance and cost risks and opportunities.
These goals can and will be reassessed and possibly revised at various stages of the project, after the R\&D phase, at placement of first orders, and during production. 
In particular, the LCF design facilitates an increase of the number of initially installed cryomodules.
The design and production schedule of LCF offers enough flexibility for possible adjustments, which will not necessarily be for the worse, with minimal impact on schedule.
The resulting cost impact is proportional to the assumed gradient change and is assessed to be no more than \SI{10}{\%} of the SRF system cost, which for LCF250LP would translate to \SI{4}{\%} of the overall project cost.
Although not included in the cost uncertainty definition adopted by ILC and LCF, this cost risk is small compared the uncertainty for the SRF system (\SI{27}{\%}) and the overall project (\SI{28}{\%}).
It does neither constitute a crippling project risk, nor is it a one-sided risk, but it is rather balanced by cost opportunities of comparable size.




}

\section*{References}
\printbibliography[heading=none]

\end{document}